\documentclass[12pt]{article}
\usepackage{color,graphicx,amssymb,amsmath,amsfonts,booktabs}

\newcommand{\mybold}[1]{{\mbox{\bf\boldmath ${#1}$}}}
\newcommand{\myboldsm}[1]{{\mbox{\scriptsize\bf\boldmath ${#1}$}}}
\newcommand{\bfc}{{\mybold{c}}}
\newcommand{\bfsmc}{{\myboldsm{c}}}
\newcommand{\bfe}{{\mybold{e}}}
\newcommand{\bfj}{{\mybold j}}

\newcommand{\calL}{{\mathcal L}}
\newcommand{\calO}{{\mathcal O}}
\newcommand{\bfr}{{\mybold{r}}}
\newcommand{\bfS}{{\mybold{S}}}
\newcommand{\bfT}{{\mybold{T}}}
\newcommand{\bfU}{{\mybold{U}}}
\newcommand{\bfsmr}{{\myboldsm{r}}}
\newcommand{\bfmu}{{\mybold{\mu}}}
\newcommand{\bfeta}{{\mybold{\eta}}}
\newcommand{\bfetasm}{{\myboldsm{\eta}}}
\newcommand{\calP}{{\mathcal P}}
\newcommand{\bfrho}{{\mybold{\rho}}}

\newcommand{\bfnabla}{{\mybold{\nabla}}}

\newcommand{\bfpi}{{\mybold{\pi}}}
\newcommand{\bfpism}{{\myboldsm{\pi}}}

\newcommand{\feq}{f^{(\mbox{\tiny eq})}}
\newcommand{\xieq}{\xi^{(\mbox{\tiny eq})}}
\newcommand{\tensor}[3]{{#1}_{#2}^{\phantom{#2}{#3}}}

\newcommand{\slantfrac}[2]{{\frac{#1}{#2}}}
\newcommand{\myvsp}{\phantom{\int_{\int_0^1}^{\int_0^1}}}

\newtheorem{example}{Example}

\begin{document}

\title{Exact Hydrodynamics of the Lattice BGK Equation
}

\author{
Bruce M. Boghosian\\
{\scriptsize Department of Mathematics, Tufts University, Medford, Massachusetts 02420, USA}\\
{\scriptsize \copyright 2008, all rights reserved}}

\date{\today}

\maketitle

\begin{abstract}
We apply the projection operator formalism to the problem of determining the asymptotic behavior of the lattice BGK equation in the hydrodynamic limit.  As an alternative to the more usual Chapman-Enskog expansion, this approach offers many benefits.  Most remarkably, it produces absolutely exact, though non-Markovian, hydrodynamic difference equations as an intermediate step.  These are accurate to all orders in Knudsen number and hence contain all of the physics of the Burnett equations and beyond.  If appropriate, these equations may then be Taylor expanded to second order in Knudsen number to obtain the usual hydrodynamic equations that result from the Chapman-Enskog analysis.  The method offers the potential to derive hydrodynamic difference equations for complex fluids with sharp gradients, such as immiscible and amphiphilic flow, for which the assumptions underlying the Chapman-Enskog approach are generally invalid.
\end{abstract}


\section{Introduction}

Statistical physics is often concerned with the problem of determining a closed set of equations of motion for a relatively small number of macroscopic degrees of freedom, given those for a much larger number of underlying degrees of freedom.  This problem occurs in the classical theoretical context of deriving fluid dynamic equations from underlying kinetic equations~\cite{bib:chapmancowling}, and also in the more modern computational context of so-called {\it multiphysics} simulations~\cite{bib:multiphysics1,bib:multiphysics2}.  These examples span a wide range of difficulty.

The derivation of the Boltzmann equation assumes excellent separation between three different scales of length and time:
\begin{enumerate}
\item The scales associated with a molecular collision are assumed to be the smallest and fastest of all the relevant scales.  In particular, the range of intermolecular force is assumed to be much smaller than a mean-free path, and the duration of a collision is assumed to be much smaller than a mean-free time.
\item The scales associated with the spatial and temporal intervals between collisions -- the mean-free path and the mean-free time, respectively -- are assumed to be much larger than the collision duration, but smaller than any hydrodynamic length and time scales.
\item Hydrodynamic length and time scales are assumed to be longest and slowest.
\end{enumerate}
Under these assumptions, an asymptotic approach may be adopted.  This is the basis of the Chapman-Enskog expansion~\cite{bib:chapmancowling} with which it is possible to derive the Navier-Stokes equations of viscous hydrodynamics.

Over the past few decades, it has been recognized that a breakdown of scale separation between scales 1 and 2 is much less catastrophic than a breakdown between scales 2 and 3.  In a dense gas or liquid, for example, the mean-free path may be comparable to the interaction range, so good separation is lost between scales 1 and 2.  This invalidates Boltzmann's {\it stosszahlansatz}, by which the Boltzmann equation is derived.  In spite of this, as long as there remains good separation between scales 2 and 3, kinetic ring theory shows that the form of the Navier-Stokes equations is robust, and the interparticle correlations introduced by the loss of separation between scales 1 and 2 may be accounted for by an appropriate renormalization of the hydrodynamic transport coefficients~\cite{bib:kineticring}.  The Navier-Stokes equations hold reasonably well for water at standard temperature and pressure, after all, even though scales 1 and 2 are comparable.

A breakdown of separation between scales 2 and 3 is a much more serious issue.  The ratio of mean-free path to hydrodynamic scale lengths is called the Knudsen number, Kn, and the Chapman-Enskog analysis is asymptotic in this quantity.  Even for single-component, single-phase fluids, the Navier-Stokes equations may break down spectacularly when this quantity becomes too large.  For complex fluid configurations, such as immiscible flow or two-phase coexistence, spatial gradients of order parameters may be very large indeed.  The characteristic width of the interface between two immiscible fluids, for example, may be on the order of a mean-free path, so that the corresponding Knudsen number is of order unity.  In this circumstance, asymptotic methods are not a viable option.

For the last decade, physicists, chemists, applied mathematicians and engineers faced with the problem of modeling complex fluids have studied lattice models of hydrodynamics.  These models consist of particle populations moving about on a lattice and colliding at lattice sites, whose emergent hydrodynamic behavior is that of a Navier-Stokes fluid~\cite{bib:lb}.  It was found much easier to introduce effective forces between particle populations on a lattice than to introduce such forces in a continuum setting~\cite{bib:shanchen}.  In this way, the dynamics of immiscible~\cite{bib:immiscible}, coexisting~\cite{bib:coexisting}, and amphiphilic~\cite{bib:amphiphilic} fluids have all been modeled successfully.

It may be argued that the success of lattice models of fluids is purely phenomenological in nature.  Attractive or repulsive interparticle potentials are introduced to make immiscible species separate, and surface tension is emergent.  Potentials with both attractive and repulsive regions, in the spirit of the Van der Waals potential, are introduced, and the liquid-gas coexistence is emergent.  Simplified BGK-style collision operators are used~\cite{bib:bgk}.  As long as the dimensionless parameters associated with the simulation match the fluid being modeled, the details of the microscopic interactions are deemed unimportant.

Faith in this phenomenological approach is, to some extent, justified by the observed robustness of hydrodynamic equations to details of the kinetic interactions.  If both the interparticle potential range and the mean-free path are on the order of a lattice spacing, loss of separation between scales 1 and 2 is evident.  As noted above, however, the hydrodynamic equations for a simple fluid are robust with respect to this loss; the hope is that a similar thing is true for more complex fluids.

Another reason for faith in the lattice-BGK approach owes to its second-order accuracy.  Mathematically, the lattice-BGK equation may be written
\begin{equation}
f_j\left(\bfr+\bfc_j,t+1\right) = 
f_j(\bfr,t) + \frac{1}{\tau}\left[\feq_j\left(\bfrho(\bfr,t)\right)-f_j(\bfr,t)\right],
\label{eq:bgk}
\end{equation}
where $f_j(\bfr,t)$ is the discrete-velocity distribution function corresponding to the $j$th velocity $\bfc_j$ at spatial position $\bfr$ and time $t$, likewise $\feq_j$ is an equilibrium distribution function that depends only on the conserved densities $\bfrho$, and $\tau$ is a collisional relaxation time.  Second-order accuracy is not at all obvious from a cursory inspection of Eq.~(\ref{eq:bgk}).  To make it more evident, Dellar~\cite{bib:dellar} has pointed out that we may define the new dependent variable
\begin{equation}
F_j :=
f_j-\frac{1}{2\tau}\left(\feq_j-f_j\right).
\end{equation}
It is then straightforward to derive the lattice BGK equation for the transformed variable.  Using $P$ to denote the {\it propagation operator}, defined by
\begin{equation}
Pf_j(x,t) = f_j(x+\bfc_j,t+1),
\end{equation}
the result may be written
\begin{equation}
PF_j = F_j +
\frac{1}{\tau-\frac{1}{2}}\left(\frac{I+P}{2}\right)\left[\feq_j-f_j\right],
\label{eq:PIform}
\end{equation}
where $I$ is the identity operator. This form makes manifest the fact that the collision is applied between sites.  It also allows a glimpse at the origin of the term $\tau-1/2$ which will emerge as a factor in the transport coefficient.

One might hope that the success of lattice BGK models would represent progress toward the end of deriving hydrodynamic equations for complex fluids from first principles.  For example, it would be very satisfying if a Chapman-Enskog analysis of an interacting-particle lattice-BGK equation gave rise to a Ginzburg-Landau or Cahn-Hilliard equation for phase separation, coupled to the Navier-Stokes equations in the style of Halperin and Hohenberg's Model H~\cite{bib:hh}.  To date, however, a successful derivation of this sort does not exist, most likely due to the aforementioned loss of separation between scales 2 and 3 for such fluids.  There is simply no small parameter analogous to the Knudsen number on which to base an asymptotic expansion.

Part of the problem is that the very first step of the Chapman-Enskog analysis of the lattice-BGK equation is the Taylor expansion of the propagation operator, effectively in powers of the Knudsen number.  This has the effect of yielding partial differential equations (PDEs) in space and time.  In this work, we argue that the insistence on PDEs is the real problem.  If we were willing to entertain the possibility of hydrodynamic equations that are difference equations in space and/or time, we would be able to model fluids with rapidly varying hydrodynamic fields.  The question arises:  Is it possible to derive a closed-form equation for the hydrodynamic degrees of freedom of a lattice BGK model without Taylor expanding, or doing anything else tantamount to an asymptotic expansion in Knudsen number?  In this paper, we answer this question affirmatively.

We circumvent the usual need for Taylor expansion by applying the projection operator formalism to the problem of deriving the exact hydrodynamics of the lattice-BGK equation.  As an alternative to the more usual Chapman-Enskog expansion, this approach offers many benefits.  Most remarkably, it produces absolutely exact, though non-Markovian, hydrodynamic difference equations as an intermediate step.  These are accurate to all orders in Knudsen number and hence contain all of the physics of the Burnett equations and beyond for the lattice BGK fluid.  If appropriate -- but {\it only} if appropriate -- these equations may then be Taylor expanded in space and/or time (to second order in Knudsen number) to obtain the hydrodynamic equations that would have resulted from the Chapman-Enskog analysis.  The method thereby offers the potential to derive hydrodynamic difference equations for complex fluids with sharp gradients, such as immiscible and amphiphilic flow, for which the assumptions underlying the Chapman-Enskog approach are generally invalid.

We begin by applying the methodology to a lattice-BGK model for Burgers' equation.  This example is simple enough to display every step of the calculation, but complicated enough to raise most all of the above issues.  In particular, spatially varying initial conditions may lead to a shock of characteristic width equal to the square root of the viscosity.  For reasonable values of the viscosity, this shock width may be on the order of a lattice spacing.

\section{Burgers' equation}
\subsection{Lattice BGK model}

The method is best illustrated by example, so we begin by applying it to a lattice BGK model for Burgers' equation in one spatial dimension ($D=1$).  A number of such models for Burgers' equation have been developed over the years~\cite{bib:lgburgers,bib:lbburgers,bib:elboburgers}; this one is a variant of an entropic version considered by Boghosian et al.~\cite{bib:elboburgers}.  At each point $x\in{\mathbb Z}$, and at each time $t\in{\mathbb Z}^+$, we have a two-component distribution function $f_\pm(x,t)$, from which it is possible to recover the hydrodynamic density
\begin{equation}
\rho(x,t) = f_+(x,t) + f_-(x,t).
\label{eq:forward}
\end{equation}
The distribution function obeys the lattice BGK kinetic equation
\begin{equation}
f_\pm\left(x\pm 1,t+1\right) = 
f_\pm(x,t) + \frac{1}{\tau}\left[\feq_\pm\left(\rho(x,t)\right)-f_\pm(x,t)\right],
\label{eq:bgkBurgers}
\end{equation}
where we have defined the local equilibrium distribution function
\begin{equation}
\feq_\pm\left(\rho\right) := \frac{\rho}{2} \pm \frac{\kappa}{2}\rho\left(1-\frac{\rho}{2}\right).
\end{equation}
Here $\kappa$ is a parameter that is taken to be of first order in the scaling limit.  That is, as the number of lattice points per physical distance is doubled, $\kappa$ will be halved.

For the analysis of this model, we shall also find it useful to define the kinetic moment
\begin{equation}
\xi(x,t) := f_+(x,t) - f_-(x,t),
\end{equation}
so that the distribution function may be recovered from knowledge of its hydrodynamic and kinetic moments,
\begin{equation}
f_\pm(x,t) = \frac{\rho(x,t) \pm \xi(x,t)}{2}.
\label{eq:backward}
\end{equation}

\subsection{Chapman-Enskog analysis}

The simple lattice BGK model described above has pedagogical value as a simple introduction to the Chapman-Enskog asymptotic procedure.  To make this presentation self-contained, and to facilitate comparison of the proposed new methodology with the Chapman-Enskog procedure, we outline the latter in this subsection.

The usual analysis begins with a Taylor expansion of the kinetic equation (\ref{eq:bgkBurgers}) assuming parabolic ordering, wherein spatial derivatives are first order quantities, and time derivatives are second order quantities.  The result is
\begin{equation}
\epsilon^2 \partial_t f_\pm \pm
\epsilon \partial_x f_\pm + \frac{\epsilon^2}{2}\partial_x^2 f_\pm +
\calO\left(\epsilon^3\right)
 = 
\frac{1}{\tau}\left[\frac{\rho}{2} \pm \frac{\epsilon\kappa}{2}\rho\left(1-\frac{\rho}{2}\right) -
f_\pm\right],
\label{eq:bgkBurgersOrd}
\end{equation}
where all quantities are understood to be evaluated at $(x,t)$, and where $\epsilon$ is a formal expansion parameter, introduced for purposes of bookkeeping, that will be set to one at the end of the calculation.  Note that we made the substitution $\kappa\rightarrow\epsilon\kappa$ to reflect the fact that $\kappa$ is first-order in the scaling limit.

Taking the sum of the $\pm$ components of this equation yields the conservation equation
\begin{equation}
\epsilon^2 \partial_t\rho+\epsilon\partial_x\xi +
\frac{\epsilon^2}{2}\partial_x^2 \rho +
\calO\left(\epsilon^3\right) = 0.
\label{eq:hydro}
\end{equation}
To close this equation, it will be necessary to express $\xi$ in terms of $\rho$.  This is done by solving Eq.~(\ref{eq:bgkBurgersOrd}) perturbatively by taking
\begin{equation}
f_\pm = f^{(0)}_\pm + \epsilon f^{(1)}_\pm + \epsilon^2 f^{(2)}_\pm +
\calO\left(\epsilon^3\right).
\end{equation}

At order zero in $\epsilon$, we immediately obtain
\begin{equation}
f^{(0)}_\pm = \frac{\rho}{2}.
\end{equation}
This lowest approximation to $f_\pm$ implies $\xi=0$; it is therefore insufficient to calculate the kinetic moment, which first appears at order one.

Proceeding to order one, we obtain
\begin{equation}
\pm \partial_x f^{(0)}_\pm =
\frac{1}{\tau}\left[\pm \frac{\kappa}{2}\rho\left(1-\frac{\rho}{2}\right) -
f^{(1)}_\pm\right],
\end{equation}
so
\begin{equation}
f^{(1)}_\pm = \pm \frac{\kappa}{2}\rho\left(1-\frac{\rho}{2}\right)\mp
\frac{\tau}{2}\partial_x \rho,
\end{equation}
where we have used the order zero solution in the last step.  This order-one result for the distribution function,
\begin{equation}
f_\pm =
\frac{\rho}{2}\pm
\frac{\epsilon\kappa}{2}\rho\left(1-\frac{\rho}{2}\right)\mp
\frac{\epsilon\tau}{2}\partial_x \rho +
\calO\left(\epsilon^2\right)
\end{equation}
yields the kinetic moment
\begin{equation}
\xi =
\epsilon\kappa\rho\left(1-\frac{\rho}{2}\right)-
\epsilon\tau\partial_x \rho +
\calO\left(\epsilon^2\right).
\end{equation}
Inserting this result in Eq.~(\ref{eq:hydro}) yields
\begin{equation}
\epsilon^2 \partial_t\rho+\epsilon\partial_x
\left[\epsilon\kappa\rho\left(1-\frac{\rho}{2}\right)-
\epsilon\tau\partial_x \rho\right] +
\frac{\epsilon^2}{2}\partial_x^2 \rho +
\calO\left(\epsilon^3\right) = 0,
\label{eq:burgers1}
\end{equation}
or
\begin{equation}
\partial_t\rho+\kappa\left(1-\rho\right)\partial_x\rho =
\left(\tau-\frac{1}{2}\right)\partial^2_x \rho +
\calO\left(\epsilon\right).
\label{eq:burgers2}
\end{equation}
Upon the substitution $u:=\kappa\left(1-\rho\right)$, this becomes Burgers' equation in canonical form,
\begin{equation}
\partial_t u+u\,\partial_x u =
\nu\,\partial^2_x u +
\calO\left(\epsilon\right),
\end{equation}
where we have defined the transport coefficient $\nu := \tau - 1/2$.

\begin{figure}
\centering
\mbox{
\includegraphics[bbllx=0,bblly=0,bburx=300,bbury=210,width=2.0truein]{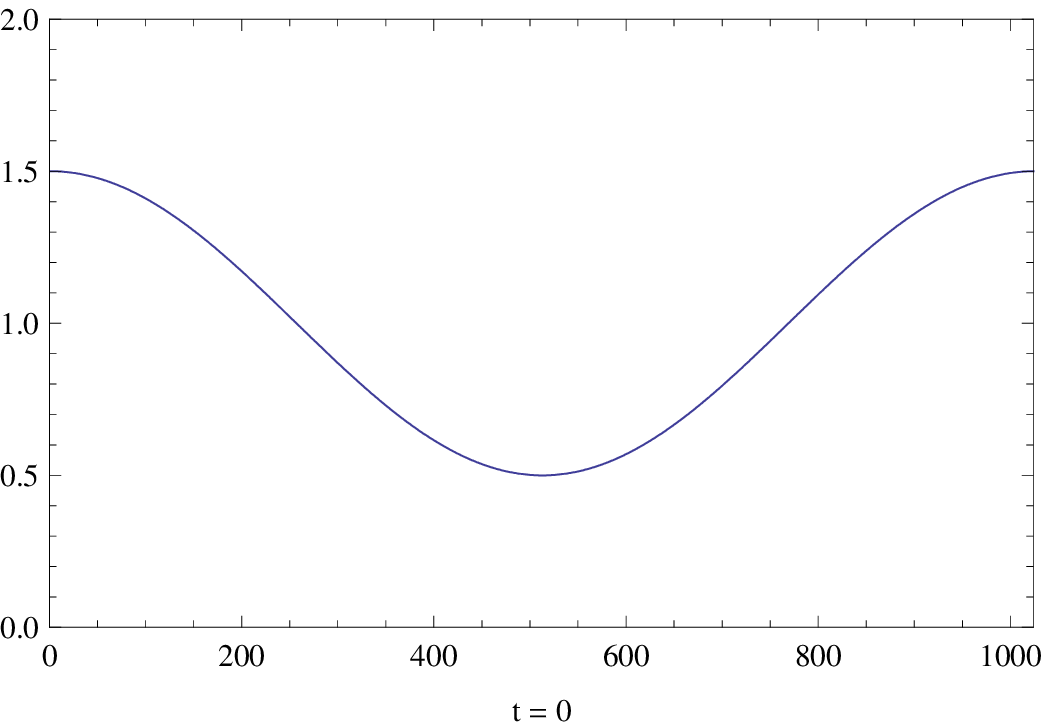}
\includegraphics[bbllx=0,bblly=0,bburx=300,bbury=210,width=2.0truein]{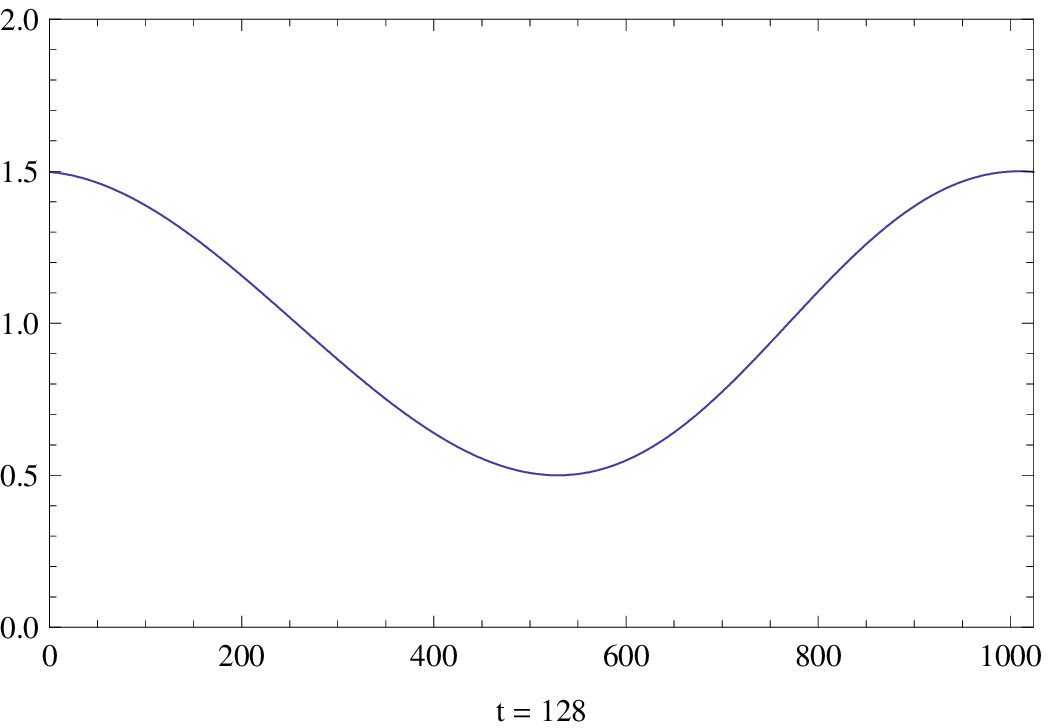}}
\mbox{
\includegraphics[bbllx=0,bblly=0,bburx=300,bbury=210,width=2.0truein]{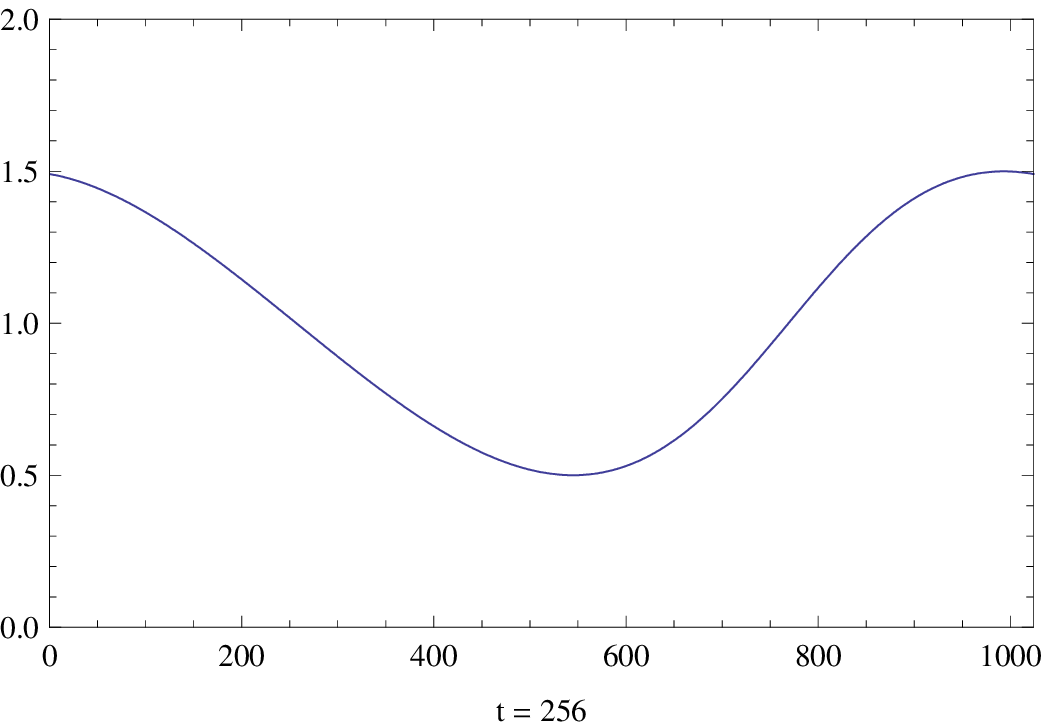}
\includegraphics[bbllx=0,bblly=0,bburx=300,bbury=210,width=2.0truein]{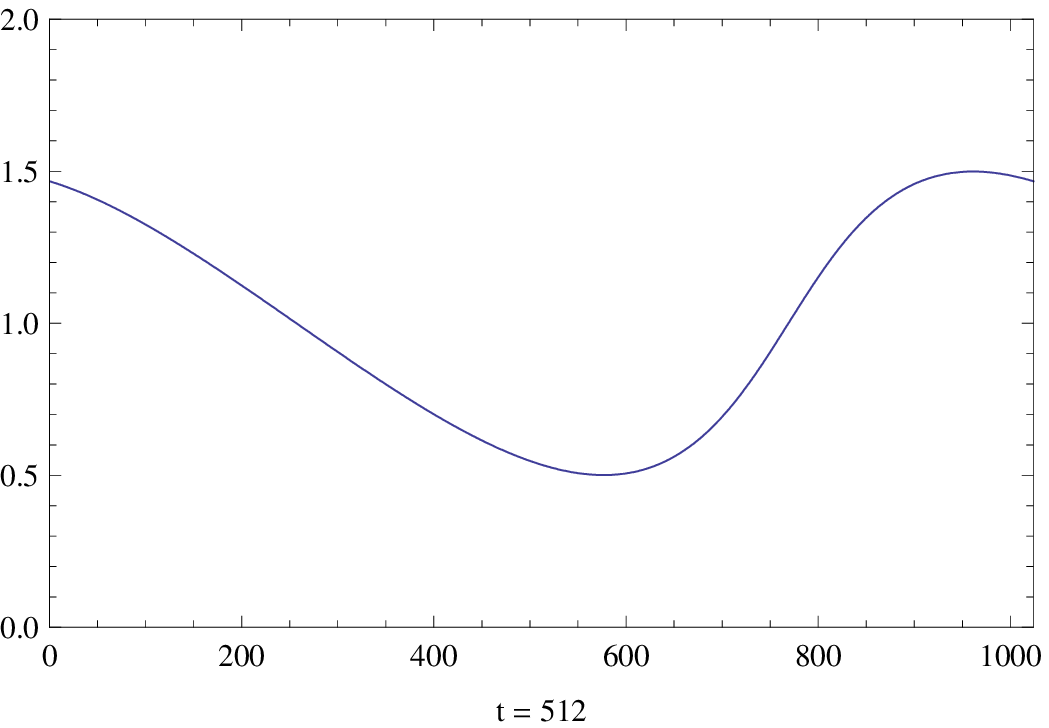}}
\mbox{
\includegraphics[bbllx=0,bblly=0,bburx=300,bbury=210,width=2.0truein]{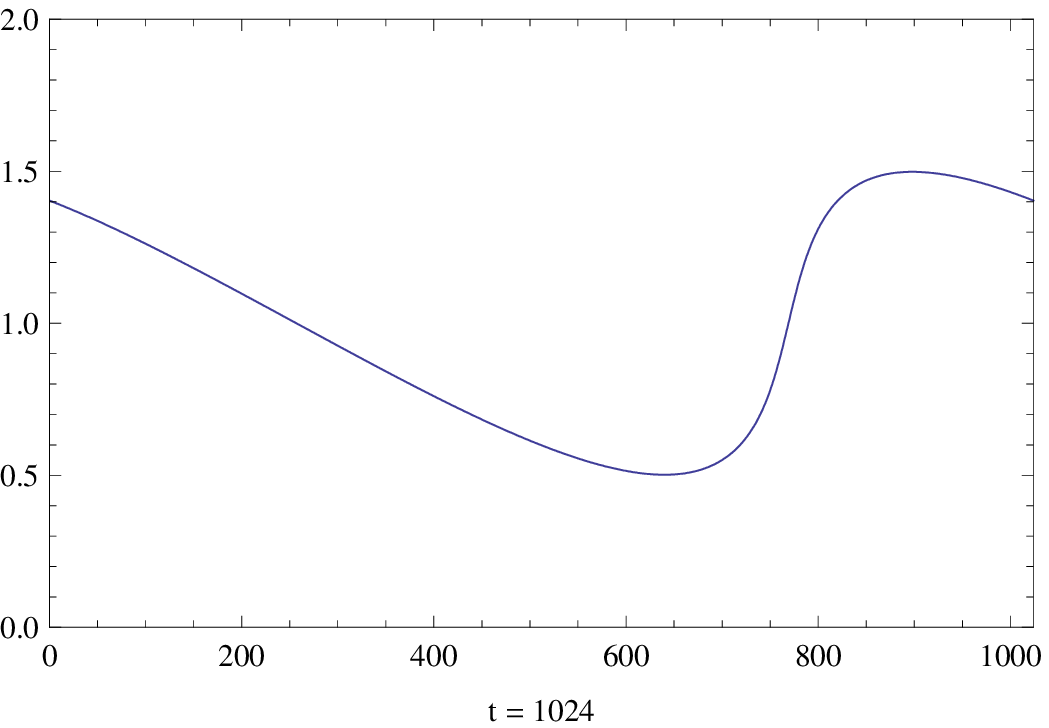}
\includegraphics[bbllx=0,bblly=0,bburx=300,bbury=210,width=2.0truein]{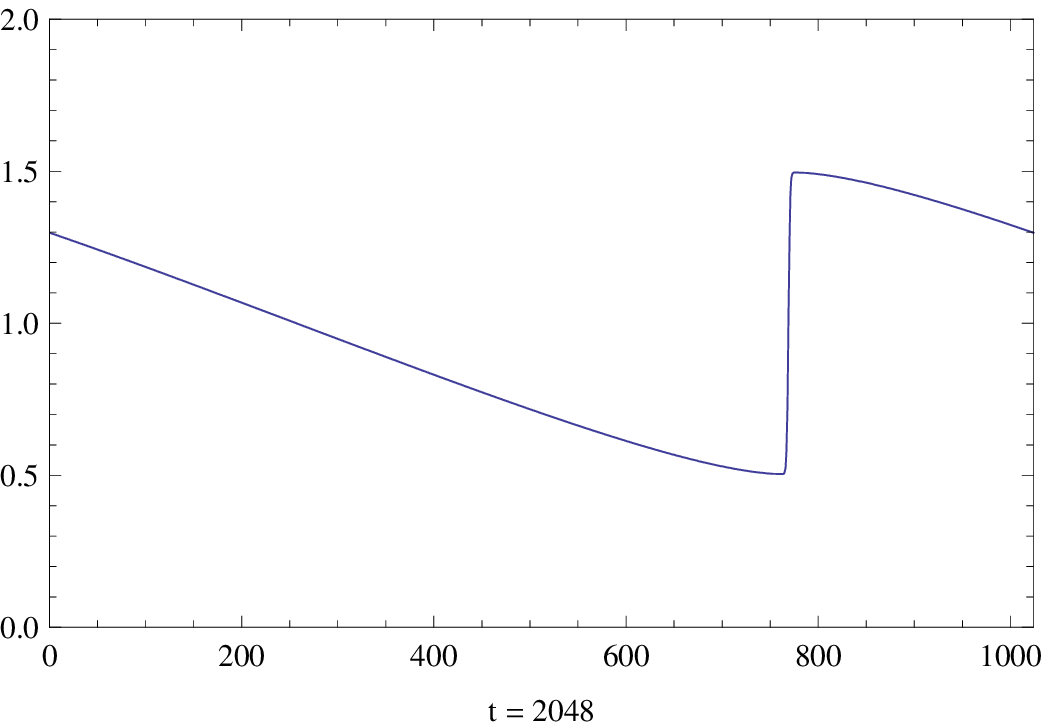}}
\mbox{
\includegraphics[bbllx=0,bblly=0,bburx=300,bbury=210,width=2.0truein]{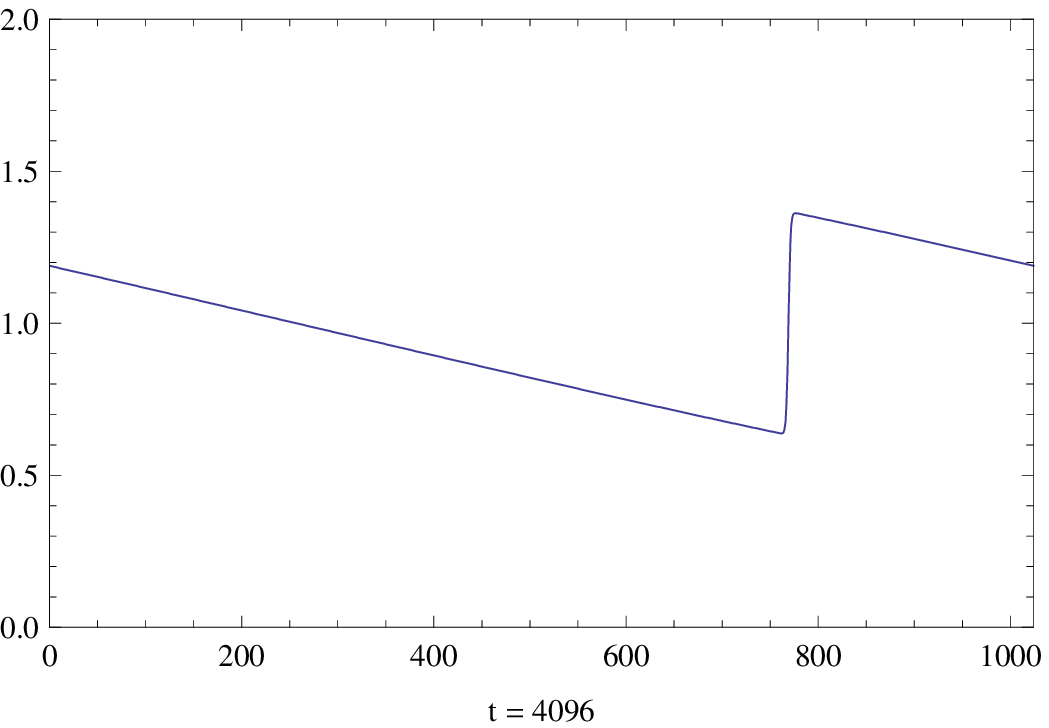}
\includegraphics[bbllx=0,bblly=0,bburx=300,bbury=210,width=2.0truein]{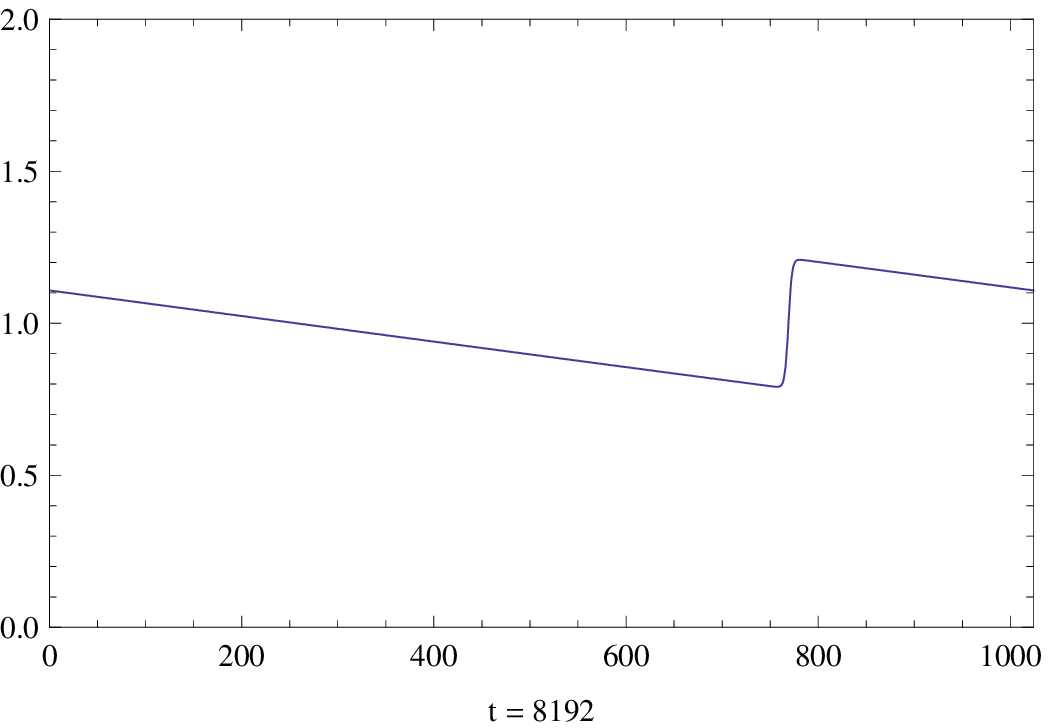}}
\caption{Simulation of lattice BGK model for Burgers' equation.  The formation and subsequent decay of the shock are evident.  Note that the shock width is on the order of a lattice spacing, calling into question the very first step of the Chapman-Enskog analysis, namely the assumption that spatial gradients are small.}
\label{fig:burgers}
\end{figure}

This method of simulating Burgers' equation is simple to implement and remarkably robust.  Fig.~\ref{fig:burgers} shows the results of simulating of the model on a periodic lattice of size $N=1024$, with $\kappa=0.25$ and $\tau=0.60$.  The initial conditions used were
\begin{equation}
\rho(x,0) = 1+\frac{1}{2}\cos\left(\frac{2\pi x}{N}\right).
\end{equation}
It is seen that the solution captures the formation and decay of the shock, and that the width of the shock at late times is comparable to the lattice spacing.

Before going on to the exact analysis, it is worth making some general observations about the Chapman-Enskog analysis for this model:
\begin{itemize}
\item We solved the kinetic equation only to first order, but used that solution in the hydrodynamic equation at second order.  This interlacing of orders is characteristic of the Chapman-Enskog analysis.
\item The second-order accuracy of the lattice BGK equation is not at all evident in the final result obtained.  If we were to carry on to the next order -- which would involve solving the kinetic equation to second order -- it is not at all clear that we would not find corrections to the hydrodynamic equation obtained.
\item The transport coefficient is equal to $\tau-1/2$.  The first term, $\tau$, arose from the gradient correction to the local equilibrium distribution function.  The second term, $-1/2$, came from the Taylor expansion of the left-hand side of the kinetic equation at order two, and is an artifact of the discrete nature of the model.  Note that the combination $\tau-1/2$ is manifest in Eq.~(\ref{eq:PIform}).
\end{itemize}

\subsection{Exact analysis}

The exact analysis that is the point of this paper comes from projecting the kinetic equation onto hydrodynamic and kinetic subspaces, solving for the kinetic field $\xi(x,t)$ as though the hydrodynamic field $\rho(x,t)$ were a known function of position and time, and then using this solution to obtain a closed dynamical equation for $\rho(x,t)$.  This general technique has been known for a long time under a variety of different names.  In physics, it is sometimes called the Mori-Zwanzig projection operator formalism~\cite{bib:zwanzig}.  In linear algebra, it is related to the Schur complement~\cite{bib:schur}. To the best of our knowledge, however, it has not heretofore been applied to the problem of obtaining exact solutions of the lattice BGK equation.

We begin by using Eqs.~(\ref{eq:forward},\ref{eq:bgkBurgers},\ref{eq:backward}) to write coupled evolution equations for $\rho$ and $\xi$.  After a bit of straightforward algebra, we obtain
\begin{eqnarray}
\lefteqn{\rho(x,t+1)
=
+\frac{1}{2}\left[\rho(x+1,t)+\rho(x-1,t)\right]-
\frac{\tau-1}{2\tau}\left[\xi(x+1,t)-\xi(x-1,t)\right]}\nonumber\\
& & \;\;\;\;\;\;\;\;\;\;\;\;\;\; +
\frac{\kappa}{2\tau}
\left[
-\rho(x+1,t)
\left(
1-\frac{\rho(x+1,t)}{2}
\right) +
\rho(x-1,t)
\left(
1-\frac{\rho(x-1,t)}{2}
\right)
\right]\label{eq:rhoProj}
\\
\lefteqn{\xi(x,t+1)
=
-\frac{1}{2}\left[\rho(x+1,t)-\rho(x-1,t)\right]+
\frac{\tau-1}{2\tau}\left[\xi(x+1,t)+\xi(x-1,t)\right]}\nonumber\\
& & \;\;\;\;\;\;\;\;\;\;\;\;\;\; +
\frac{\kappa}{2\tau}
\left[
+\rho(x+1,t)
\left(
1-\frac{\rho(x+1,t)}{2}
\right) +
\rho(x-1,t)
\left(
1-\frac{\rho(x-1,t)}{2}
\right)
\right].\label{eq:piProj}
\end{eqnarray}
These two dynamical equations for the hydrodynamic and kinetic fields respectively, taken together, are equivalent to the original kinetic equation.  The equation for the kinetic field may be written in the suggestive form
\begin{equation}
\xi(x,t+1) =
\frac{\tau-1}{2\tau}\left[\xi(x+1,t)+\xi(x-1,t)\right]+Q(x,t),
\label{eq:dhe}
\end{equation}
where we have grouped together everything involving the hydrodynamic field into a single effective source term, defined by
\begin{eqnarray}
Q(x,t) 
&:=&
-\frac{1}{2}
\left[\rho(x+1,t)-\rho(x-1,t)\right]\nonumber\\
& &
+\frac{\kappa}{2\tau}
\left[
\rho(x+1,t)\left(1-\frac{\rho(x+1,t)}{2}\right)+
\rho(x-1,t)\left(1-\frac{\rho(x-1,t)}{2}\right)
\right].
\label{eq:qdef}
\end{eqnarray}

Eq.~(\ref{eq:dhe}) is a linear difference equation.  For an infinite lattice, it has the exact solution
\begin{equation}
\xi(x,t) =
\sum_{n=0}^{t-1}\sum_{m=0}^n\binom{n}{m}
\left(\frac{\sigma}{2}\right)^n Q(x+2m-n,t-n) +
\sum_{m=0}^{t}\binom{t}{m}
\left(\frac{\sigma}{2}\right)^{t} \xi(x+2m-t,0),
\label{eq:xiSol}
\end{equation}
where we have defined $\sigma :=1-1/\tau$, and we note that $|\sigma|<1$ for $\tau>1/2$.  The first term above describes the excitation of the kinetic field due to gradients of the hydrodynamic field.  The second term is a transient, due to the initial conditions.

\subsection{Self-consistent hydrodynamic difference equation}

Supposing, for the time being, that the kinetic field is initialized to zero~\footnote{It should be evident that this is an inessential assumption that we make here only for the sake of simplicity of presentation.}, or that we have waited long enough for transient behavior to be unimportant, we can insert the first term on the right-hand side of Eq.~(\ref{eq:xiSol}) into the dynamical equation for $\rho$ and rearrange to obtain the nonlinear difference equation
\begin{eqnarray}
\lefteqn{\rho(x,t+1)-\rho(x,t)
=
+\frac{1}{2}\left[\rho(x+1,t)-2\rho(x,t)+\rho(x-1,t)\right]}\nonumber\\
& &
-\frac{\kappa}{2\tau}\left\{
\rho(x+1,t)\left[1-\frac{\rho(x+1,t)}{2}\right] -
\rho(x-1,t)\left[1-\frac{\rho(x-1,t)}{2}\right]
\right\}\nonumber\\
& &
+\frac{1}{2}\sum_{n=0}^{t-1}\sum_{m=0}^n\binom{n}{m}
\left(\frac{\sigma}{2}\right)^{n+1}\nonumber\\
& &
\phantom{\int\int}
\left[\rho(x+2m-n+2,t-n)-2\rho(x+2m-n,t-n)+\rho(x+2m-n-2,t-n)\right]
\nonumber\\
& &
-\frac{\kappa}{2\tau}\sum_{n=0}^{t-1}\sum_{m=0}^n\binom{n}{m}
\left(\frac{\sigma}{2}\right)^{n+1}
\left\{
\rho(x+2m-n+2,t-n)\left[1-\frac{\rho(x+2m-n+2,t-n)}{2}\right]
\right.\nonumber\\
& &
\phantom{\int\int}
\left.
-\rho(x+2m-n-2,t-n)\left[1-\frac{\rho(x+2m-n-2,t-n)}{2}\right]
\right\}.
\label{eq:exact1}
\end{eqnarray}
The remarkable thing about this result is that no essential approximations have been made in its derivation.  It is an exact difference equation that must be obeyed by a hydrodynamic field $\rho(x,t)$ satisfying the lattice BGK equation.

\subsection{Recovery of a hydrodynamic differential equation}

Examination of Fig.~\ref{fig:burgers} indicates that the hydrodynamic field changes reasonably slowly in time so that a Taylor expansion in time is justified, but that it can change very rapidly in space after the onset of the shock, rendering a spatial Taylor expansion of questionable validity.  For this reason, we begin by Taylor expanding in time only, incurring an error of at most second order,
\begin{eqnarray}
\lefteqn{
\partial_t\rho(x,t)
=
+\frac{1}{2}\left[\rho(x+1,t)-2\rho(x,t)+\rho(x-1,t)\right]}\nonumber\\
& &
-\frac{\kappa}{2\tau}\left\{
\rho(x+1,t)\left[1-\frac{\rho(x+1,t)}{2}\right] -
\rho(x-1,t)\left[1-\frac{\rho(x-1,t)}{2}\right]
\right\}\nonumber\\
& &
+\frac{1}{2}\sum_{n=0}^{t-1}\sum_{m=0}^n\binom{n}{m}
\left(\frac{\sigma}{2}\right)^{n+1}\nonumber\\
& &
\phantom{\int\int}
\left[\rho(x+2m-n+2,t)-2\rho(x+2m-n,t)+\rho(x+2m-n-2,t)\right]
\nonumber\\
& &
-\frac{\kappa}{2\tau}\sum_{n=0}^{t-1}\sum_{m=0}^n\binom{n}{m}
\left(\frac{\sigma}{2}\right)^{n+1}
\left\{
\rho(x+2m-n+2,t)\left[1-\frac{\rho(x+2m-n+2,t)}{2}\right]
\right.\nonumber\\
& &
\phantom{\int\int}
\left.
-\rho(x+2m-n-2,t)\left[1-\frac{\rho(x+2m-n-2,t)}{2}\right]
\right\}+\calO\left(\epsilon^2\right).
\label{eq:exact2}
\end{eqnarray}
This is a second-order accurate set of coupled discrete-space, continuous-time equations describing the hydrodynamics of the model.

To proceed to a spatiotemporal partial differential equation, we may Taylor expand Eq.~(\ref{eq:exact2}) about the spatial point $x$, retaining spatial derivatives to second order.  Because of the symmetry of the differences, it is straightforward to see that the error incurred is of second order.  The sum over $m$ is then a binomial series and that over $n$ is a finite geometric series, yielding
\begin{equation}
\partial_t\rho+
\kappa\left(1-\sigma^{t+1}\right)
\left(1-\rho\right)\partial_x\rho =
\left[\tau\left(1-\sigma^{t+1}\right)-\frac{1}{2}\right]\partial_x^2\rho +
\calO\left(\epsilon^2\right).
\end{equation}
Ignoring the terms that decay exponentially in time (note $|\sigma|<1$), we recover Eq.~(\ref{eq:burgers2}).

Before going on to consider the exact analysis for more general lattice BGK equations, we offer some preliminary observations:
\begin{itemize}
\item The usual Chapman-Enskog procedure begins with a Taylor series expansion of the kinetic equation in difference form, and then assembles continuum hydrodynamic equations from that series.  This method, by contrast, first derives exact hydrodynamic equations in difference form, and only then (optionally) Taylor expands the results to obtain differential equations.
\item Closely related to the above point is the idea that the hydrodynamic difference equation first obtained is not expanded in Knudsen number, and therefore expected to hold in limits that would not ordinarily be considered ``hydrodynamic.''  Such limits may include situations with steep spatial gradients or long mean-free paths.
\item The method yields an exact hydrodynamic equation in difference form that can be Taylor expanded to yield the hydrodynamic equation.  The second-order accurate nature of this expansion is manifest.
\item The method works only for lattice BGK equations, and relies on the fact that the equilibrium distribution function is a function of the conserved quantities only.  (If this were not the case, Eq.~(\ref{eq:dhe}) would not be linear in the kinetic field, so we would not be able to solve it exactly.)
\item Note that as $t\rightarrow\infty$, the effect of the last two terms on the right-hand side of Eq.~(\ref{eq:exact2}) is to simply alter the coefficients in front of -- or ``renormalize'' -- the other terms in the equation.  The penultimate term evaluates to $C\rho_{xx}$ so it renormalizes the diffusion term, and the last term is $-D\kappa\rho(1-\rho)\rho_x$ so it renormalizes the advection term.  The sum determines that $C=\sigma\tau$ and $D=\sigma$ whence
\begin{equation}
\partial_t\rho = \frac{1}{2}\rho_{xx}
-\frac{1}{\tau}\partial_x\left[\kappa\rho\left(1-\frac{\rho}{2}\right)\right]
+\sigma\tau\rho_{xx}
-\sigma\partial_x\left[\kappa\rho\left(1-\frac{\rho}{2}\right)\right],
\end{equation}
from which Eq.~(\ref{eq:burgers2}) follows.
\end{itemize}

\section{General procedure}

\subsection{Projection operators}

We now suppose that we have a $b$-component discrete-velocity distribution function $f_j(\bfr,t)$, where $j\in\{1,\ldots,b\}$, where $\bfr\in\calL$ is a point on lattice $\calL$, and where $t\in{\mathbb Z}^+$ is the time.  The $b_h$ hydrodynamic variables are obtained by projection with the $b_h\times b$ matrix $H$,
\begin{eqnarray}
\rho_\alpha &=& \sum_j^b \tensor{H}{\alpha}{j} f_j,\\
\noalign{\noindent \mbox{and the $b_k=b-b_h$ kinetic variables are obtained by projection with the $b_k\times b$ matrix $K$,}}
\xi_\mu &=& \sum_j^b \tensor{K}{\mu}{j} f_j.
\end{eqnarray}
The rows of $H$ and $K$ are linearly independent, so that knowledge of all the hydrodynamic and kinetic variables is sufficient to reconstruct the distribution function,
\begin{equation}
f_j = \sum_\beta^{b_h} \tensor{A}{j}{\beta}\rho_\beta + \sum_\nu^{b_k} \tensor{B}{j}{\nu}\xi_\nu.
\label{eq:reconstruct}
\end{equation}
It is easily verified that the $b\times b_h$ matrix $A$ and the $b\times b_k$ matrix $B$ obey the relations
\begin{eqnarray}
\sum_j^b \tensor{H}{\alpha}{j}\tensor{A}{j}{\beta} =
\tensor{\delta}{\alpha}{\beta}
& &
\sum_j^b \tensor{H}{\alpha}{j}\tensor{B}{j}{\nu} = 0
\label{eq:ida}\\
\sum_j^b \tensor{K}{\mu}{j}\tensor{A}{j}{\beta} = 0
& &
\sum_j^b \tensor{K}{\mu}{j}\tensor{B}{j}{\nu} =
\tensor{\delta}{\mu}{\nu},
\end{eqnarray}
and
\begin{equation}
\tensor{\delta}{j}{i} = \sum_\beta^{b_h}
\tensor{A}{j}{\beta}\tensor{H}{\beta}{i} +
\sum_\nu^{b_k} \tensor{B}{j}{\nu}\tensor{K}{\nu}{i}.
\label{eq:idc}
\end{equation}
In the above presentation, we have adopted the convention of using Greek letters from the beginning of the alphabet ($\alpha,\beta,\ldots$) to label the hydrodynamic variables, from the middle of the alphabet ($\mu,\nu,\ldots$) to label the kinetic variables, and Latin letters ($j,k,\ldots$) to label the distribution function components.  We shall adhere to this notational convention as closely as possible in the forthcoming development.

\begin{example}
\label{ex:triangular}
A triangular grid in two spatial dimensions ($D=2$) has the $b=6$ lattice vectors
\begin{equation}
\bfc_j := \bfe_x\cos\left(\frac{2\pi j}{6}\right) + \bfe_y\sin\left(\frac{2\pi j}{6}\right).
\end{equation}
In the widely adopted nomenclature for lattice Boltzmann models, this is called the D2Q6 model~\footnote{We note in passing that the D2Q6 model is no longer extensively used for two-dimensional lattice BGK simulations of fluids.  It has been abandoned in favor of the so-called D2Q9 lattice which may be implemented on a Cartesian grid.  We use the D2Q6 model in this example only for the sake of simplicity, since it allows us to exhibit matrices of dimension six rather than nine.  It should be clear that there is nothing preventing this method from being applied to the D2Q9 lattice, or even to the much larger lattices used in modern lattice-BGK simulations.}.  If mass and momentum are conserved, the hydrodynamic and kinetic projection operators may be taken to be
\begin{equation}
H =
\left[
\begin{array}{rrrrrr}
1 & 1 & 1 & 1 & 1 & 1\\
1 & \slantfrac{1}{2} & -\slantfrac{1}{2} & -1 & -\slantfrac{1}{2} & \slantfrac{1}{2}\\
0 & \slantfrac{\sqrt{3}}{2} & \slantfrac{\sqrt{3}}{2} & 0 & -\slantfrac{\sqrt{3}}{2} & -\slantfrac{\sqrt{3}}{2}
\end{array}
\right]\phantom{,}
\label{eq:H}
\end{equation}
and
\begin{equation}
K =
\left[
\begin{array}{rrrrrr}
1 & -1 & 1 & -1 & 1 & -1\\
1 & -\slantfrac{1}{2} & -\slantfrac{1}{2} & 1 & -\slantfrac{1}{2} & -\slantfrac{1}{2}\\
0 & \slantfrac{\sqrt{3}}{2} & -\slantfrac{\sqrt{3}}{2} & 0 & \slantfrac{\sqrt{3}}{2} & -\slantfrac{\sqrt{3}}{2}
\end{array}
\right],
\end{equation}
respectively.  Note that the first row of $H$ corresponds to the mass density $\rho$, while the second and third rows correspond to the momentum density $\bfpi$.  Collectively, we refer to the conserved densities as $\bfrho=(\rho,\bfpi)$.  The corresponding reconstruction matrices are then
\begin{equation}
\begin{array}{ccc}
A =
\left[
\begin{array}{rrr}
\slantfrac{1}{6} & \slantfrac{1}{3} & 0\\
\slantfrac{1}{6} & \slantfrac{1}{6} & \slantfrac{\sqrt{3}}{6}\\
\slantfrac{1}{6} & -\slantfrac{1}{6} & \slantfrac{\sqrt{3}}{6}\\
\slantfrac{1}{6} & -\slantfrac{1}{3} & 0\\
\slantfrac{1}{6} & -\slantfrac{1}{6} & -\slantfrac{\sqrt{3}}{6}\\
\slantfrac{1}{6} & \slantfrac{1}{6} & -\slantfrac{\sqrt{3}}{6}\\
\end{array}
\right] & &
B =
\left[
\begin{array}{rrr}
\slantfrac{1}{6} & \slantfrac{1}{3} & 0\\
-\slantfrac{1}{6} & -\slantfrac{1}{6} & \slantfrac{\sqrt{3}}{6}\\
\slantfrac{1}{6} & -\slantfrac{1}{6} & -\slantfrac{\sqrt{3}}{6}\\
-\slantfrac{1}{6} & \slantfrac{1}{3} & 0\\
\slantfrac{1}{6} & -\slantfrac{1}{6} & \slantfrac{\sqrt{3}}{6}\\
-\slantfrac{1}{6} & -\slantfrac{1}{6} & -\slantfrac{\sqrt{3}}{6}\\
\end{array}
\right]
\end{array}
\label{eq:AB}
\end{equation}
The identities of Eqs.~(\ref{eq:ida}) through (\ref{eq:idc}) are then readily verified.
\label{ex:tri}
\end{example}

\subsection{General form of the lattice BGK equation}

The general form of the lattice BGK equation, Eq.~(\ref{eq:bgk}), may be rewritten
\begin{equation}
f_j(\bfr,t+1) = \sigma f_j(\bfr-\bfc_j,t) + \left(1-\sigma\right)
\feq_j\left(\bfrho\left(\bfr-\bfc_j,t\right)\right),
\end{equation}
where $\sigma := 1-1/\tau$ is defined as before.  Note that the equilibrium distribution function $\feq$ is allowed to depend only on the hydrodynamic moments $\bfrho$.  In what follows, it shall prove useful to write the equilibrium distribution in the form
\begin{equation}
\feq_j(\bfrho)=\sum_\beta^{b_h}\tensor{A}{j}{\beta}\rho_\beta + \xieq_j(\bfrho).
\end{equation}
Comparing this with Eq.~(\ref{eq:reconstruct}), we see that $\xieq_j(\bfrho)$ is the kinetic portion of the equilibrium distribution function.
\begin{example}
For a fluid with $\bfrho=(\rho,\bfpi)$, the form generally used for this dependence is the Mach-expanded distribution
\begin{equation}
\feq_j(\bfrho) =
W_j\left[\rho +
\frac{1}{c_s^2}\bfpi\cdot\bfc_j +
\frac{1}{2c_s^4\rho}\bfpi\bfpi :
\left(\bfc_j\bfc_j - c_s^2 I_2\right)
\right],
\label{eq:feq}
\end{equation}
where the $W_j$ are weights associated with each direction, $I_2$ is the rank-two unit tensor and $c_s$ is the sound speed defined by the isotropy requirement
\begin{equation}
\sum_j^b W_j\bfc_j\bfc_j = c_s^2 I_2\sum_j^b W_j.
\end{equation}
In particular, for the D2Q6 lattice of Example~\ref{ex:tri}, it may be verified that $W_j=1$ and $c_s=1/\sqrt{2}$.  The first two terms of Eq.~(\ref{eq:feq}) are then the hydrodynamic portion of the equilibrium distribution function, and
\begin{equation}
\xieq_j(\bfrho) = 
\frac{W_j}{2c_s^4\rho}\bfpi\bfpi :
\left(\bfc_j\bfc_j - c_s^2 I_2\right)
\end{equation}
is the kinetic portion.
\end{example}

By means of the projection operators defined in the previous subsection, it is straightforward to decompose this into coupled evolution equations for the hydrodynamic and kinetic moments,
\begin{eqnarray}
\rho_\alpha(\bfr,t+1)
&=&
\sigma \sum_j^b \tensor{H}{\alpha}{j}
\left[
\sum_\beta^{b_h} \tensor{A}{j}{\beta}\rho_\beta(\bfr-\bfc_j,t) +
\sum_\nu^{b_k} \tensor{B}{j}{\nu}\xi_\nu(\bfr-\bfc_j,t)\right]\nonumber\\
& &
+ \left(1-\sigma\right)\sum_j^b
\tensor{H}{\alpha}{j}\,\feq_j\left(\bfrho\left(\bfr-\bfc_j,t\right)\right)
\label{eq:rhoGen}\\
\noalign{\noindent \mbox{and}}
\xi_\mu(\bfr,t+1)
&=&
\sigma \sum_j^b \tensor{K}{\mu}{j}
\left[
\sum_\beta^{b_h} \tensor{A}{j}{\beta}\rho_\beta(\bfr-\bfc_j,t) +
\sum_\nu^{b_k} \tensor{B}{j}{\nu}\xi_\nu(\bfr-\bfc_j,t)\right]\nonumber\\
& &
+ \left(1-\sigma\right)\sum_j^b
\tensor{K}{\mu}{j}\,\feq_j\left(\bfrho\left(\bfr-\bfc_j,t\right)\right),
\label{eq:piGen}
\end{eqnarray}
respectively.  Eqs.~(\ref{eq:rhoGen}) and (\ref{eq:piGen}) are the generalizations of Eqs.~(\ref{eq:rhoProj}) and (\ref{eq:piProj}), respectively.  The equation for the kinetic modes may be written more succinctly as
\begin{equation}
\xi_\mu(\bfr,t+1)
=
\sigma \sum_{\nu}^{b_k}\sum_j^b \left(
\tensor{K}{\mu}{j}\tensor{B}{j}{\nu}\right)\xi_\nu(\bfr-\bfc_j,t) +
Q_\mu(\bfr,t),
\label{eq:kinGen}
\end{equation}
where we have defined
\begin{equation}
Q_\mu(\bfr,t) :=
\sum_{\beta}^{b_h}\sum_j^b \left(\tensor{K}{\mu}{j}\tensor{A}{j}{\beta}\right)\rho_\beta(\bfr-\bfc_j,t)
+ \left(1-\sigma\right)\sum_j^b
\tensor{K}{\mu}{j}\,\xieq_j\left(\bfrho\left(\bfr-\bfc_j,t\right)\right).
\label{eq:qdefg}
\end{equation}
Eqs.~(\ref{eq:kinGen}) and (\ref{eq:qdefg}) are the generalizations of Eqs.~(\ref{eq:dhe}) and (\ref{eq:qdef}), respectively.

\subsection{Exact analysis in the general case}

To proceed as in the example, it is now necessary to find an exact solution to the linear equation Eq.~(\ref{eq:kinGen}) for the kinetic modes, assuming that the hydrodynamic modes are known.  

\begin{figure}
\centering
\mbox{\includegraphics[bbllx=0,bblly=100,bburx=1330,bbury=1147,width=6.0truein]{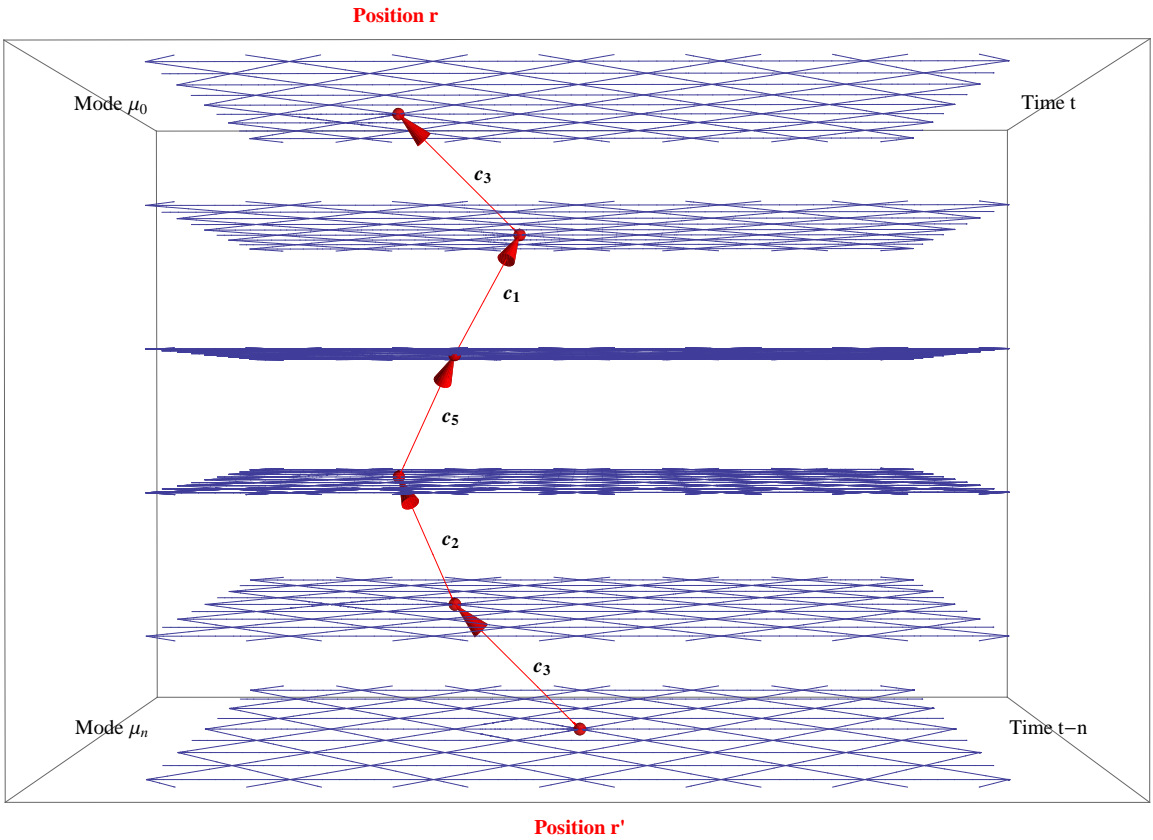}}
\caption{Path from $\bfr'$ to $\bfr$ in the D2Q6 model}
\label{fig:path}
\end{figure}

Consider a labeled path of $n$ steps along lattice vectors, whose vertices are labeled by kinetic modes.  More specifically, consider a path along the lattice beginning at position $\bfr'$ and mode $\mu_n=\nu$ at time $t'=t-n$, and ending at position $\bfr\in\calL$ and mode $\mu_0=\mu$ at time $t$.  One such path in the D2Q6 model is illustrated in Fig.~\ref{fig:path}.  The set of all such $n$-step paths will be denoted by $\calP^{(n)}(\bfr',\nu;\bfr,\mu)$.

A path $p\in\calP^{(n)}(\bfr',\nu;\bfr,\mu)$ is thus characterized by its sequence of indices $\bfj(p):=\{j_n,\ldots,j_1\}$ of the lattice vectors traversed, and also the sequence of kinetic modes $\bfmu:=\{\mu_n,\ldots,\mu_0\}$ at the visited vertices.  Note that it must be true that
\begin{equation}
\sum_{\ell=1}^n \bfc_{j_\ell} = \bfr-\bfr',
\label{eq:addrule}
\end{equation}
and
\begin{eqnarray}
\mu_0&=&\mu\\
\mu_n&=&\nu.
\end{eqnarray}
We use $\Sigma_p^{\calP^{(n)}(\bfr',\nu;\bfr,\mu)}$ to denote the sum over all paths, $p\in\calP^{(n)}(\bfr',\nu;\bfr,\mu)$.

\begin{example}
A path of length $n=5$ in the D2Q6 model of Example~\ref{ex:triangular} is shown in Fig.~\ref{fig:path}.  The sequence of lattice vector indices pictured is $\bfj(p)=\{3,2,5,1,3\}$.
\end{example}

To a path $p$ we assign the {\it weight}
\begin{equation}
\tensor{w^{(n)}}{\mu}{\nu}\left(\bfj,\bfmu\right) = \prod_{\ell=1}^n
\left(\tensor{K}{\mu_{\ell-1}}{j_\ell}\tensor{B}{j_\ell}{\mu_{\ell}}\right),
\end{equation}
where there is no understood summation on repeated indices.  The exact solution to Eq.~(\ref{eq:kinGen}) is then
\begin{eqnarray}
\xi_\mu(\bfr,t)
&=&
\sum_{n=0}^{t-1}
\sum_{\bfsmr'\in\calL}
\sum_\nu^{b_k}
\sum_{p}^{\calP^{(n)}(\bfsmr',\nu;\bfsmr,\mu)}
\sigma^n\;
\tensor{w^{(n)}}{\mu}{\nu}\left(\bfj(p),\bfmu(p)\right)\;
Q_\nu(\bfr',t-n)\nonumber\\
& &
+
\sum_{\bfsmr'\in\calL}
\sum_\nu^{b_k}
\sum_{p}^{\calP^{(t)}(\bfsmr',\nu;\bfsmr,\mu)}
\sigma^{t}\;
\tensor{w^{(t)}}{\mu}{\nu}\left(\bfj(p),\bfmu(p)\right)\;
\xi_\nu(\bfr',0).
\label{eq:xiSolGen}
\end{eqnarray}
Note that if $\bfr'$ can not be connected to $\bfr$ by a sequence of $n$ lattice vectors, perhaps because it is too far away, then $\calP^{(t)}(\bfr',\mu_t;\bfr,\mu_0)$ is understood to be the null set.

\begin{example}
Eq.~(\ref{eq:xiSolGen}) is the generalization of Eq.~(\ref{eq:xiSol}).  To see this, note that our $D=1$ example for Burgers' equation had $b_h=b_k=1$ and $b=2$.  The projection operators were
\begin{eqnarray}
H
&=&
\left[
\begin{array}{cc}
+1 & +1
\end{array}
\right]\\
K
&=&
\left[
\begin{array}{cc}
+1 & -1
\end{array}
\right],
\end{eqnarray}
and
\begin{eqnarray}
A
&=&
\left[
\begin{array}{c}
+1/2\\
+1/2
\end{array}
\right]\\
B
&=&
\left[
\begin{array}{c}
+1/2\\
-1/2
\end{array}
\right],
\end{eqnarray}
from which it follows that $\tensor{w^{(n)}(\bfj,\bfmu)}{1}{1}=2^{-n}$.  The number of paths connecting $\bfr'=x+2m-n$ and $\bfr=x$ is then the binomial coefficient $\binom{n}{m}$, resulting in Eq.~(\ref{eq:xiSol}).
\end{example}

\subsection{Self-consistent hydrodynamic difference equations}

As with our one-dimensional example, we suppose that the kinetic field is initialized to zero, and we insert Eq.~(\ref{eq:xiSolGen}) into Eq.~(\ref{eq:rhoGen}) and rearrange to obtain the {\it exact} hydrodynamic difference equation
\begin{eqnarray}
\lefteqn{\rho_\alpha(\bfr,t+1)}\nonumber\\
&=&
\phantom{+}\sigma \sum_j^b
\sum_\beta^{b_h} \tensor{H}{\alpha}{j} \tensor{A}{j}{\beta}\rho_\beta(\bfr-\bfc_j,t)
+ \left(1-\sigma\right)\sum_j^b
\tensor{H}{\alpha}{j}\,\feq_j\left(\bfrho\left(\bfr-\bfc_j,t\right)\right)
\nonumber\\
& &
+\sum_j^b
\sum_{\mu,\nu}^{b_k}\sum_{n=0}^{t-1}\sum_{\bfsmr'\in\calL}\;
\sum_{p}^{\calP^{(n)}(\bfsmr',\nu;\bfsmr-\bfsmc_j,\mu)}
\sigma^{n+1}\;
\tensor{H}{\alpha}{j} \tensor{B}{j}{\mu}
\tensor{w^{(n)}}{\mu}{\nu}\left(\bfj(p),\bfmu(p)\right)\;
Q_\nu(\bfr',t-n).
\label{eq:rhoGenSol}
\end{eqnarray}

Eq.~(\ref{eq:rhoGenSol}) is an exact nonlinear hydrodynamic difference equation for the conserved densities, albeit in terms of a diagrammatic summation.  We are now free to Taylor expand in either time or space, as appropriate to the phenomenon under consideration.  Taylor expansion of Eq.~(\ref{eq:rhoGenSol}) to second order in the time variable followed by a bit of rearranging yields the continuum-time, discrete-space hydrodynamic equation
\begin{eqnarray}
\lefteqn{\partial_t\rho_\alpha(\bfr,t)}\nonumber\\
&=&
\phantom{+}\sum_j^b
\sum_\beta^{b_h} \tensor{H}{\alpha}{j} \tensor{A}{j}{\beta}
\left[\rho_\beta(\bfr-\bfc_j,t) - \rho_\beta(\bfr,t)\right]
\nonumber\\
& &
+ \left(1-\sigma\right)\sum_j^b
\tensor{H}{\alpha}{j}\,
\left[\xieq_j\left(\bfrho\left(\bfr-\bfc_j,t\right)\right) -
\xieq_j\left(\bfrho\left(\bfr,t\right)\right)\right]
\nonumber\\
& &
+\sum_j^b
\sum_{\mu,\nu}^{b_k}\sum_{n=0}^{t-1}\sum_{\bfsmr'\in\calL}\;
\sum_{p}^{\calP^{(n)}(\bfsmr',\nu;\bfsmr-\bfsmc_j,\mu)}
\sigma^{n+1}\;
\tensor{H}{\alpha}{j} \tensor{B}{j}{\mu}
\tensor{w^{(n)}}{\mu}{\nu}\left(\bfj(p),\bfmu(p)\right)\;
Q_\nu(\bfr',t).
\label{eq:rhoGenSol2}
\end{eqnarray}

\subsection{Recovery of a hydrodynamic differential equation}

Finally, to proceed to a hydrodynamic differential equation, we may Taylor expand in space, retaining terms to second order.  While this step is not technically difficult, it is tedious enough to warrant splitting the calculation into three parts, corresponding to each of the three terms on the right-hand side of Eq.~(\ref{eq:rhoGenSol2}).  We begin by defining some useful tensors in terms of the projection and reconstruction operators and the lattice vectors.

\subsubsection{Some useful tensors}

In preparation for the forthcoming analysis, it is useful to define the tensors\begin{eqnarray}
\tensor{S}{\alpha}{\beta}(N) &:=&
\sum_j^b\tensor{H}{\alpha}{j} \tensor{A}{j}{\beta}\bigotimes^N\bfc_j\\
\tensor{T}{\alpha}{\beta}(N) &:=&
\sum_j^b\tensor{H}{\alpha}{j} \frac{\partial\xi_j}{\partial\rho_\beta}\bigotimes^N\bfc_j\\
\tensor{U}{\alpha}{\beta\gamma}(N) &:=&
\sum_j^b\tensor{H}{\alpha}{j} \frac{\partial^2\xi_j}{\partial\rho_\beta\partial\rho_\gamma}\bigotimes^N\bfc_j,
\label{eq:uNdef}
\end{eqnarray}
where $\otimes^N$ denotes an $N$-fold outer product.  If all of the conserved quantities are scalars, these tensors have $N$ spatial indices ranging from $1$ to $D$ in which they are completely symmetric; that is, all have spatial rank $N$.  In addition, the $\bfS(N)$ and $\bfT(N)$ each have two hydrodynamic indices ranging, $\alpha$ and $\beta$, ranging from $1$ to $b_h$, and the $\bfU(N)$ have three such hydrodynamic indices.

When it becomes necessary to refer to these tensors by their spatial indices, we adopt the convention of replacing the number $N$ in parentheses by the actual list of $N$ spatial indices.  Thus, for example, we have
\begin{equation}
\tensor{S}{\alpha}{\beta}(\{a,b,c\}) =
\sum_j^b\tensor{H}{\alpha}{j} \tensor{A}{j}{\beta}c_{ja}c_{jb}c_{jc},
\end{equation}
where $c_{ja}$ is the $a$th spatial component of lattice vector $\bfc_j$.  Note that the value of $N$, namely $N=3$ in this case, can be inferred by the fact that there are three indices in the list.  When listing components for $N=0$, we denote the empty list by $\{\}$.

If one or more of the conserved quantities are spatial vectors, such as momentum, it is convenient to abuse notation by allowing the hydrodynamic indices to be either scalars or vectors.  Each one that is a vector will increase the spatial rank of the tensors by one.  For example, if $\zeta$ is the index of a conserved scalar and $\bfeta$ is the index of a conserved vector, $\tensor{S}{\zeta}{\zeta}(N)$ is a completely symmetric tensor of spatial rank $N$, while $\tensor{S}{\bfetasm}{\zeta}(N)$ is a tensor of spatial rank $N+1$ that is symmetric under interchange of $N$ of its indices, and $\tensor{S}{\bfetasm}{\bfetasm}(N)$ is a tensor of spatial rank $N+2$ that is also symmetric under interchange of $N$ of its indices.

To specify components of these, it may be necessary to use nested subscripts.  Thus, we have that
\begin{equation}
\tensor{U}{\eta_a}{\zeta\zeta}(\{b,c\}) =
\sum_j^b\tensor{H}{\eta_a}{j} \frac{\partial^2\xi_j}{\partial\rho_\zeta\partial\rho_\zeta}c_{jb}c_{jc}
\end{equation}
is the $(a,b,c)$ component of a tensor of spatial rank three that is symmetric under interchange of its rightmost two indices.  Note that the notation automatically ensures that the $N$ symmetric indices will be the rightmost indices.

Finally, we also define alternative versions of the $\bfS(N)$, $\bfT(N)$ and $\bfU(N)$ tensors, using the same names for economy of notation, but with upper hydrodynamic and lower kinetic indices,
\begin{eqnarray}
\tensor{S}{\mu}{\beta}(N) &:=&
\sum_j^b\tensor{K}{\mu}{j} \tensor{A}{j}{\beta}\bigotimes^N\bfc_j\\
\tensor{T}{\mu}{\beta}(N) &:=&
\sum_j^b\tensor{K}{\mu}{j} \frac{\partial\xi_j}{\partial\rho_\beta}\bigotimes^N\bfc_j\\
\tensor{U}{\mu}{\beta\gamma}(N) &:=&
\sum_j^b\tensor{K}{\mu}{j} \frac{\partial^2\xi_j}{\partial\rho_\beta\partial\rho_\gamma}\bigotimes^N\bfc_j.
\end{eqnarray}
All of the above-mentioned considerations about counting independent components likewise apply to these alternative versions.

\begin{example}
For the D2Q6 model in Example~\ref{ex:triangular}, supposing that mass and momentum are both conserved, the matrices $H$, $K$, $A$ and $B$ are given in Eqs.~(\ref{eq:H}) through (\ref{eq:AB}).  If we denote~\footnote{Unfortunately, denoting mass by index $\rho$ and momentum by vector index $\bfpi$ violates our convention that indices representing conserved quantities should come from the beginning of the Greek alphabet.  The standard use of these letters to represent mass and momentum density, however, was deemed to override this concern.} the hydrodynamic indices by $\alpha,\beta,\gamma\in\{\rho,\bfpi\}$, it is straightforward to compute the above tensors.  Tabulations of the results for $S(N)$, $T(N)$ and $U(N)$ are shown in Tables~\ref{eq:sH}, \ref{eq:tH} and \ref{eq:uH}, respectively.  We have not bothered listing entries that are anisotropic and/or not needed for the derivation of the hydrodynamic equations.
\end{example}

To assemble the hydrodynamic equations from Eq.~(\ref{eq:rhoGenSol2}), we label the terms on its right-hand side as \textcircled{1}, \textcircled{2} and \textcircled{3}, and consider each separately.

\subsubsection{Evaluation of first term in Eq.~(\ref{eq:rhoGenSol2})}

The first term on the right of Eq.~(\ref{eq:rhoGenSol2}) can be written
\begin{eqnarray}
\mbox{\textcircled{1}}_\alpha
&:=&
\sum_j^b
\sum_\beta^{b_h} \tensor{H}{\alpha}{j} \tensor{A}{j}{\beta}
\left[\rho_\beta(\bfr-\bfc_j,t) -
\rho_\beta(\bfr,t)\right]\nonumber\\
&=&
\sum_j^b
\sum_\beta^{b_h} \tensor{H}{\alpha}{j} \tensor{A}{j}{\beta}
\left[-\bfc_j\cdot\bfnabla\rho_\beta(\bfr,t)
+\frac{1}{2}\bfc_j\bfc_j : \bfnabla\bfnabla\rho_\beta(\bfr,t)
+\cdots\right]\nonumber\\
&=&
-\sum_\beta^{b_h}\tensor{S}{\alpha}{\beta}(1)\cdot\bfnabla\rho_\beta(\bfr,t) +
\frac{1}{2}\sum_\beta^{b_h}\tensor{S}{\alpha}{\beta}(2) : \bfnabla\bfnabla\rho_\beta(\bfr,t)+\cdots
\label{eq:FirstTerm}
\end{eqnarray}
In Appendix~\ref{sec:Afirst}, we evaluate this expression for the D2Q6 model.  When doing so, we adopt incompressible scaling~\cite{bib:LandauLifshitzFluids}, wherein spatial gradients and Mach number are taken to be order $\epsilon$, and time derivatives and density fluctuations are taken to be order $\epsilon^2$.  To leading order, we find that the $\rho$ and $\bfpi$ components of the above term are
\begin{eqnarray}
\mbox{\textcircled{1}}_\rho
&=& -\bfnabla\cdot\bfpi\\
\noalign{\noindent \mbox{and}}
\mbox{\textcircled{1}}_\bfpism
&=& -\frac{1}{2}\bfnabla\rho+\frac{1}{8}\nabla^2\bfpi.
\end{eqnarray}

\subsubsection{Evaluation of second term in Eq.~(\ref{eq:rhoGenSol2})}

Similar considerations can be applied to the second term on the right of Eq.~(\ref{eq:rhoGenSol2}) which may be written
\begin{eqnarray}
\mbox{\textcircled{2}}_\alpha
&:=&
\left(1-\sigma\right)
\sum_j^b\tensor{H}{\alpha}{j}
\left[\xieq_j(\bfrho(\bfr-\bfc_j,t)) -\xieq_j(\bfrho(\bfr,t))\right]\nonumber\\
&=&
\left(1-\sigma\right)
\sum_j^b\tensor{H}{\alpha}{j}
\left[-\bfc_j\cdot\bfnabla\xieq_j(\bfrho(\bfr,t))
+\frac{1}{2}\bfc_j\bfc_j : \bfnabla\bfnabla\xieq_j(\bfrho(\bfr,t))
+\cdots\right],
\end{eqnarray}
or
\begin{equation}
\frac{\mbox{\textcircled{2}}_\alpha}{1-\sigma}
=
-\sum_\beta^{b_h}\tensor{T}{\alpha}{\beta}(1)\cdot\bfnabla\rho_\beta
+\frac{1}{2}\sum_\beta^{b_h}\tensor{T}{\alpha}{\beta}(2) : \bfnabla\bfnabla\rho_\beta
+\frac{1}{2}\sum_{\beta,\gamma}^{b_h}\tensor{U}{\alpha}{\beta\gamma}(2) : \left(\bfnabla\rho_\beta\right)\left(\bfnabla\rho_\gamma\right)+\cdots
\label{eq:SecondTerm}
\end{equation}
In Appendix~\ref{sec:Asecond}, we evaluate this for the D2Q6 model in the limit of incompressible scaling~\cite{bib:LandauLifshitzFluids}.  To leading order, we find that the $\rho$ and $\bfpi$ components of the above term are
\begin{eqnarray}
\mbox{\textcircled{2}}_\rho
&=& 0\\
\noalign{\noindent \mbox{and}}
\mbox{\textcircled{2}}_\bfpism
&=& -\frac{1}{2\rho}\left(\bfpi\cdot\bfnabla\bfpi
-\frac{1}{2}\bfnabla\left|\bfpi\right|^2\right).
\end{eqnarray}

\subsubsection{Evaluation of third term in Eq.~(\ref{eq:rhoGenSol2})}

The third term on the right-hand side of Eq.~(\ref{eq:rhoGenSol2}) involves $Q_\mu(\bfr,t)$ which is given by Eq.~(\ref{eq:qdefg}), so it is necessary to compute this first.  We note that it may be expanded to second-order accuracy to obtain
\begin{eqnarray}
Q_\mu(\bfr,t)
&=&
\sum_{\beta}^{b_h}\tensor{S}{\mu}{\beta}(0)\rho_\beta
-\sum_{\beta}^{b_h}\tensor{S}{\mu}{\beta}(1)\cdot\bfnabla\rho_\beta
+\frac{1}{2}\sum_{\beta}^{b_h}\tensor{S}{\mu}{\beta}(2):\bfnabla\bfnabla\rho_\beta+\cdots
\nonumber\\
& &
+ \left(1-\sigma\right)
\left[
\sum_j^{b}\tensor{K}{\mu}{j}\xieq_j
-\sum_{\beta}^{b_h}\tensor{T}{\mu}{\beta}(1)\cdot\bfnabla\rho_\beta
+\frac{1}{2}\sum_{\beta}^{b_h}\tensor{T}{\mu}{\beta}(2):\bfnabla\bfnabla\rho_\beta
\right.\nonumber\\
& &
\left.
+\frac{1}{2}\sum_{\beta,\gamma}^{b_h}\tensor{U}{\mu}{\beta\gamma}(2):\left(\bfnabla
\rho_\beta\right)\left(\bfnabla\rho_\gamma\right)
+\cdots
\right].
\label{eq:QTerm}
\end{eqnarray}
In Appendix~\ref{sec:AQ}, we evaluate this for the D2Q6 model in the limit of incompressible scaling~\cite{bib:LandauLifshitzFluids}.  To second order, we find that its $\rho$ and $\bfpi$ components are
\begin{eqnarray}
Q_\rho &=& \rho\\
Q_\bfpism &=& \bfpi-\frac{1}{2}\bfnabla\rho+\frac{1}{8}\nabla^2\bfpism
-\left(\frac{1-\sigma}{2\rho}\right)
\left(\bfpi\cdot\bfnabla\bfpi-\frac{1}{2}\bfnabla\left|\bfpi\right|^2
\right)
\end{eqnarray}

As noted for the example of Burgers' equation, the effect of the diagrammatic sum as $t\rightarrow\infty$ and in the continuum limit is to apply a linear operator to these terms, thereby renormalizing other terms in the equation.  The most general form for the third term in Eq.~(\ref{eq:rhoGenSol2}) is then
\begin{equation}
Q_\alpha = C_\alpha\left[
-\frac{1}{2}\bfnabla\rho+\frac{1}{8}\nabla^2\bfpi
-\left(\frac{1-\sigma}{2\rho}\right)
\left(\bfpi\cdot\bfnabla\bfpi-\frac{1}{2}\bfnabla\left|\bfpi\right|^2
\right)\right] + D_\alpha\bfnabla\rho + E_\alpha\nabla^2\bfpi,
\end{equation}
where $C_\alpha$, $D_\alpha$ and $E_\alpha$ are determined by the diagrammatic series.  The results are
\begin{equation}
C_\rho = D_\rho = E_\rho = 0
\end{equation}
and
\begin{eqnarray}
C_\bfpism &=& \frac{1+\sigma}{1-\sigma}\\
D_\bfpism &=& \frac{1}{2}\left(\frac{1+\sigma}{1-\sigma}\right)\\
E_\bfpism &=& \frac{1}{4}\left(\frac{\sigma}{1-\sigma}\right),
\end{eqnarray}
so the third term of Eq.~(\ref{eq:rhoGenSol2}) has no $\rho$ component,
\begin{equation}
\mbox{\textcircled{3}}_\rho = 0
\end{equation}
and $\bfpi$ component equal to
\begin{eqnarray}
\mbox{\textcircled{3}}_\bfpism
&=&
\frac{1+\sigma}{1-\sigma}\left[
-\frac{1}{2}\bfnabla\rho+\frac{1}{8}\nabla^2\bfpi
-\left(\frac{1-\sigma}{2\rho}\right)
\left(\bfpi\cdot\bfnabla\bfpi
-\frac{1}{2}\bfnabla\left|\bfpi\right|^2
\right)\right]\nonumber\\
& &
+ \frac{1}{2}\left(\frac{1+\sigma}{1-\sigma}\right)\bfnabla\rho
+ \frac{1}{4}\left(\frac{\sigma}{1-\sigma}\right)\nabla^2\bfpi
\end{eqnarray}

\subsubsection{Assembly of hydrodynamic equations}

Armed with these terms, we may now assemble the full hydrodynamic equations,
\begin{eqnarray}
\rho_t &=&
\mbox{\textcircled{1}}_\rho
+ \mbox{\textcircled{2}}_ \rho
+ \mbox{\textcircled{3}}_ \rho\\
\bfpi_t &=&
\mbox{\textcircled{1}}_\bfpism
+ \mbox{\textcircled{2}}_\bfpism
+ \mbox{\textcircled{3}}_\bfpism.
\end{eqnarray}
Since $\rho_t$ may be ignored in the incompressible limit, the first of these reduces to
\begin{equation}
\bfnabla\cdot\bfpi = 0,
\label{eq:ns1}
\end{equation}
while the second gives
\begin{eqnarray}
\partial_t\bfpi
&=&
\left(1+\frac{1+\sigma}{1-\sigma}\right)
\left[
-\frac{1}{2}\bfnabla\rho
+\frac{1}{8}\nabla^2\bfpi
-\frac{1-\sigma}{2\rho}
\left(\bfpi\cdot\bfnabla\bfpi-\frac{1}{2}\bfnabla\left|\bfpi\right|^2\right)
\right]\nonumber\\
& &
+ \frac{1}{2}\left(\frac{1+\sigma}{1-\sigma}\right)\bfnabla\rho
+ \frac{1}{4}\left(\frac{\sigma}{1-\sigma}\right)\nabla^2\bfpi
\end{eqnarray}
This simplifies to yield
\begin{equation}
\partial_t\bfpi
+ \frac{1}{\rho}\bfpi\cdot\bfnabla\bfpi =
-\bfnabla P
+\nu\nabla^2\bfpi,
\label{eq:ns2}
\end{equation}
where the pressure $P$ is given by the equation of state
\begin{equation}
P = \frac{\rho}{2}-\frac{\left|\bfpi\right|^2}{2\rho},
\end{equation}
and the kinematic viscosity is
\begin{equation}
\nu = \frac{1}{2}\left(\tau - \frac{1}{2}\right).
\end{equation}
Eqs.~(\ref{eq:ns1}) and (\ref{eq:ns2}) are seen to be the Navier-Stokes equations of viscous, incompressible hydrodynamics.  The expressions for the equation of state~\footnote{The equation of state includes a term that is clearly nonphysical because it depends on the  hydrodynamic velocity.  Great concern is often expressed about this term, but it is entirely misplaced and indeed reflects a serious misunderstanding of the nature of the incompressible limit, in which the form of the equation of state is entirely irrelevant and where the pressure is given by the solution to a Poisson equation.  As long as one takes care to keep the Mach number small enough so that the density fluctuation scales as its square, the method is perfectly valid for simulating incompressible flow, the form of the equation of state notwithstanding.} and the viscosity are well known for the D2Q6 fluid~\cite{bib:SucciBook}.

\section{Discussion}

Diagrammatic methods often yield new physical insights, and this one is no exception.  Since the $Q_\alpha$ are the driving terms in the linear equations for the kinetic modes $\xi_\nu$, we may think of the $Q_\alpha$ as the precise combination of hydrodynamic gradients that excite kinetic modes.  Thus excited, the kinetic modes may propagate and couple with other kinetic modes.  The diagrams trace the evolution of these kinetic excitations in space and time until they ultimately project, via the $\tensor{B}{j}{\nu}$ matrices, a contribution back to the hydrodynamic modes.  Alternatively stated, the sum over diagrams gives the Green's function of Eq.~(\ref{eq:kinGen}) which governs the evolution of the kinetic modes.

It is remarkable that the effect of these kinetic excitations is often nothing more than the renormalization of terms already present in the hydrodynamic equation.  The diagrammatic sum is necessary to obtain the correct answer for the transport coefficients (advection, diffusion and viscosity in the above examples), but it is not necessary to obtain the general form of the hydrodynamic equation.  In addition, principles of symmetry and covariance may be employed to reduce the diagrammatic sum to the determination of a few scalar quantities -- the $A$, $B$ and $C$ quantities in the above examples.

\section{Conclusions}

We have described a new method to derive hydrodynamic equations for lattice BGK fluids.  It is qualitatively different from the usual Chapman-Enskog analysis, and superior insofar as it results in absolutely exact hydrodynamic equations.  Additionally, while more demanding in terms of calculation, the method is arguably conceptually simpler than the Chapman-Enskog approach, easier to automate using symbolic mathematics software, and more transparently a description of the generation and propagation of kinetic modes in a moving fluid.

We have illustrated this new methodology, first by presenting a simple lattice-BGK model for Burgers' equation, and second by presenting a more complex lattice-BGK model for a viscous, incompressible fluid.  The final step in this method is the extraction of a diagrammatic sum, but we have shown that all that we really need is the asymptotic form of this sum for $n$ large, and also that the effect of this sum is often nothing more than the renormalization of hydrodynamic transport coefficients.

Because the method relegates the Taylor expansion of the propagation operator to an optional step at the end of the analysis, it is possible to derive hydrodynamic equations that are either discrete or continuous in space and/or time.  For the examples presented in this work, we showed the result of expanding in time but not space.  For certain phenomena in transport theory that tend to be uniform in space but rapidly changing in time, such as spontaneous micelleization, it may be more appropriate to expand in space but not time.

Future work will take up the application of this method to complex fluid phenomena, such as multiphase fluids described by the Shan-Chen model~\cite{bib:shanchen}.  It is hoped that this method will yield accurate hydrodynamic equations in the spirit of Halperin and Hohenberg's Model H~\cite{bib:hh} for such complex fluids.

\section*{Acknowledgments}

This work was partially funded by ARO award number W911NF-04-1-0334, AFOSR award number FA9550410176, and facilitated by scientific visualization equipment funded by NSF award number 0619447.  The lattice BGK model for Burgers' equation was worked out in preparation for a presentation given at the Consiglio Nazionale delle Ricerche (CNR) in Rome, Italy on 3-8 March 2008.  Part of this work was conducted while visiting the School of Engineering at Peking University in November and December of 2007, and the Physique Nonlin\'{e}aire group at the Laboratoire de Physique Statistique of the \'{E}cole Normale Sup\'{e}rieure de Paris in April and early May of 2008.  The author would like to thank Tufts University School of Arts and Sciences for his sabbatical and FRAC leave during academic year 2007-2008 when much of this work was conducted.  Finally, the author is grateful to Jonas L\"{a}tt and Scott MacLachlan for helpful discussions.

\bibliographystyle{plain.bst}
\bibliography{paper}

\appendix

\section{Appendix:  Evaluation of Eq.~(\ref{eq:FirstTerm}) for D2Q6 fluid}
\label{sec:Afirst}

For a mass and momentum-conserving fluid, we wish to compute the $\rho$ and $\bfpi$ components of Eq.~(\ref{eq:FirstTerm}).  The $\rho$ component may be written
\begin{eqnarray}
\mbox{\textcircled{1}}_\rho
&:=&
\sum_j^b
\sum_\beta^{b_h} \tensor{H}{\rho}{j} \tensor{A}{j}{\beta}
\left[\rho_\beta(\bfr-\bfc_j,t) -
\rho_\beta(\bfr,t)\right]\nonumber\\
&=&
-\tensor{S}{\rho}{\rho}(\{a\})\nabla_a\rho
-\tensor{S}{\rho}{\pi_a}(\{b\})\nabla_b\pi_a\nonumber\\
& &
+\frac{1}{2}\tensor{S}{\rho}{\rho}(\{a,b\})\nabla_a\nabla_b\rho
+\frac{1}{2}\tensor{S}{\rho}{\pi_a}(\{b,c\})\nabla_b\nabla_c\pi_a+\cdots
\end{eqnarray}
Using the results for the D2Q6 model compiled in Tables~\ref{eq:sH}, \ref{eq:tH} and \ref{eq:uH}, this becomes
\begin{eqnarray}
\mbox{\textcircled{1}}_\rho
&=&
-\delta_{ab}\nabla_b\pi_a
+\frac{1}{4}\delta_{ab}\nabla_a\nabla_b\rho+\cdots\nonumber\\
&=&
-\bfnabla\cdot\bfpi
+\frac{1}{4}\nabla^2\rho+\cdots
\end{eqnarray}
When incompressible scaling is taken into account, the second term is order $\epsilon^2$ times the first one, so we may write the result,
\begin{equation}
\boxed{
\mbox{\textcircled{1}}_\rho =
-\bfnabla\cdot\bfpi+\calO\left(\epsilon^2\right).}
\end{equation}

The $\bfpi$ component may be written
\begin{eqnarray}
\mbox{\textcircled{1}}_{\pi_a}
&:=&
\sum_j^b
\sum_\beta^{b_h} \tensor{H}{\pi_a}{j} \tensor{A}{j}{\beta}
\left[\rho_\beta(\bfr-\bfc_j,t) -
\rho_\beta(\bfr,t)\right]\nonumber\\
&=&
-\tensor{S}{\pi_a}{\rho}(\{b\})\nabla_b\rho
-\tensor{S}{\pi_a}{\pi_b}(\{c\})\nabla_c\pi_b
\end{eqnarray}
Using the results for the D2Q6 model compiled in Tables~\ref{eq:sH}, \ref{eq:tH} and \ref{eq:uH}, this becomes
\begin{eqnarray}
\mbox{\textcircled{1}}_{\pi_a}
&=&
+\frac{1}{2}\tensor{S}{\pi_a}{\rho}(\{b,c\}) \nabla_b\nabla_c\rho
+\frac{1}{2}\tensor{S}{\pi_a}{\pi_b}(\{c,d\}) \nabla_c\nabla_d\pi_b+\cdots\nonumber\\
&=&
-\frac{1}{2}\delta_{ab}\nabla_b\rho
+\frac{1}{8}\left(\delta_{ab}\delta_{cd}+\delta_{ac}\delta_{bd}+\delta_{ad}\delta_{bc}\right) \nabla_c\nabla_d\pi_b+\cdots
\end{eqnarray}
When incompressible scaling is taken into account, the second term is order $\epsilon^2$ times the first one, so we may write the result,
\begin{equation}
\boxed{
\mbox{\textcircled{1}}_{\bfpism}
=
-\frac{1}{2}\bfnabla\rho
+\frac{1}{8}\left[
\nabla^2\bfpi
+2\bfnabla\left(\bfnabla\cdot\bfpi\right)
\right] +\calO\left(\epsilon^2\right).}
\end{equation}

\section{Appendix:  Evaluation of Eq.~(\ref{eq:SecondTerm}) for D2Q6 fluid}
\label{sec:Asecond}

For a mass and momentum-conserving fluid, we wish to compute the $\rho$ and $\bfpi$ components of Eq.~(\ref{eq:SecondTerm}).  The $\rho$ component may be written
\begin{eqnarray}
\frac{\mbox{\textcircled{2}}_\rho}{1-\sigma}
&:=&
\sum_j^b\tensor{H}{\rho}{j}
\left[\xieq_j(\bfrho(\bfr-\bfc_j,t)) -\xieq_j(\bfrho(\bfr,t))\right]\nonumber\\
&=&
-\tensor{T}{\rho}{\rho}(\{a\})\nabla_a\rho
-\tensor{T}{\rho}{\pi_a}(\{b\})\nabla_b\pi_a\nonumber\\
& &
+\frac{1}{2}\tensor{T}{\rho}{\rho}(\{a,b\})\nabla_a\nabla_b\rho
+\frac{1}{2}\tensor{T}{\rho}{\pi_a}(\{b,c\})\nabla_b\nabla_c\pi_a\nonumber\\
& &
+\frac{1}{2}\tensor{U}{\rho}{\rho\rho}(\{a,b\})\left(\nabla_a\rho\right)\left(\nabla_b\rho\right)
+\frac{1}{2}\tensor{U}{\rho}{\rho\pi_a}(\{b,c\})\left(\nabla_b\rho\right)\left(\nabla_c\pi_a\right)\nonumber\\
& &
+\frac{1}{2}\tensor{U}{\rho}{\pi_a\rho}(\{b,c\})\left(\nabla_b\pi_a\right)\left(\nabla_c\rho\right)
+\frac{1}{2}\tensor{U}{\rho}{\pi_a\pi_b}(\{c,d\})\left(\nabla_c\pi_a\right)\left(\nabla_d\pi_b\right)+\cdots\nonumber\\
&=&
+\frac{1}{8\rho^2}\left(\pi^2\delta_{ab}-2\pi_a\pi_b\right)\nabla_a\nabla_b\rho
+\frac{1}{4\rho}\left(\delta_{ab}\pi_c+\delta_{ac}\pi_b-\delta_{bc}\pi_a\right)\nabla_b\nabla_c\pi_a\nonumber\\
& &
-\frac{1}{4\rho^3}\left(\pi^2\delta_{ab}-\pi_a\pi_b\right)\left(\nabla_a\rho\right)\left(\nabla_b\rho\right)
-\frac{1}{4\rho^2}\left(\delta_{ab}\pi_c+\delta_{ac}\pi_b-\delta_{bc}\pi_a\right)\left(\nabla_b\rho\right)\left(\nabla_c\pi_a\right)\nonumber\\
& &
-\frac{1}{4\rho^2}\left(\delta_{ab}\pi_c+\delta_{ac}\pi_b-\delta_{bc}\pi_a\right)\left(\nabla_b\pi_a\right)\left(\nabla_c\rho\right)\nonumber\\
& &
+\frac{1}{4\rho}\left(\delta_{ac}\delta_{bd}+\delta_{ad}\delta_{bc}-\delta_{ab}\delta_{cd}\right)\left(\nabla_c\pi_a\right)\left(\nabla_d\pi_b\right)+\cdots
\end{eqnarray}
Using the results for the D2Q6 model compiled in Tables~\ref{eq:sH}, \ref{eq:tH} and \ref{eq:uH}, this becomes
\begin{eqnarray}
\frac{\mbox{\textcircled{2}}_\rho}{1-\sigma}
&=&
\frac{1}{8\rho^2}\left(\left|\bfpi\right|^2\nabla^2\rho-2\bfpi\bfpi : \bfnabla\bfnabla\rho\right)
+\frac{1}{4\rho}\left[\bfpi\cdot\bfnabla\left(\bfnabla\cdot\bfpi\right)+\bfpi\cdot\bfnabla\left(\bfnabla\cdot\bfpi\right)-\bfpi\cdot\nabla^2\bfpi\right]\nonumber\\
& &
-\frac{1}{4\rho^3}\left[\left|\bfpi\right|^2\left|\bfnabla\rho\right|^2-\left(\bfpi\cdot\bfnabla\rho\right)^2\right]\nonumber\\
& &
-\frac{1}{4\rho^2}\left[\left(\bfpi\cdot\bfnabla\bfpi\right)\cdot\left(\bfnabla\rho\right)
+\left(\bfpi\cdot\bfnabla\rho\right)\left(\bfnabla\cdot\bfpi\right)
-\left(\bfnabla\rho\right)\cdot\left(\bfnabla\bfpi\right)\cdot\bfpi\right]\nonumber\\
& &
-\frac{1}{4\rho^2}\left[\left(\bfpi\cdot\bfnabla\rho\right)\left(\bfnabla\cdot\bfpi\right)
+\left(\bfpi\cdot\bfnabla\bfpi\right)\cdot\left(\bfnabla\rho\right)
-\frac{1}{2}\left(\bfnabla\left|\bfpi\right|^2\right)\cdot\left(\bfnabla\rho\right)\right]\nonumber\\
& &
+\frac{1}{4\rho}\left\{
\left(\bfnabla\cdot\bfpi\right)^2
+\mbox{Tr}\left[\left(\bfnabla\bfpi\right)\cdot\left(\bfnabla\bfpi\right)\right]
-\left(\bfnabla\bfpi\right):\left(\bfnabla\bfpi\right)\right\}+\cdots
\end{eqnarray}
All of these terms are negligible when incompressible scaling is applied, so we have
\begin{equation}
\boxed{
\mbox{\textcircled{2}}_\rho
=
0+\calO\left(\epsilon^2\right).}
\end{equation}

The $\bfpi$ component may be written
\begin{eqnarray}
\frac{\mbox{\textcircled{2}}_{\pi_a}}{1-\sigma}
&:=&
\sum_j^b\tensor{H}{\pi_a}{j}
\left[\xieq_j(\bfrho(\bfr-\bfc_j,t)) -\xieq_j(\bfrho(\bfr,t))\right]\nonumber\\
&=&
-\tensor{T}{\pi_a}{\rho}(\{b\})\nabla_b\rho
-\tensor{T}{\pi_a}{\pi_b}(\{c\})\nabla_c\pi_b\nonumber\\
& &
+\frac{1}{2}\tensor{T}{\pi_a}{\rho}(\{b,c\})\nabla_b\nabla_c\rho
+\frac{1}{2}\tensor{T}{\pi_a}{\pi_b}(\{c,d\})\nabla_c\nabla_d\pi_b\nonumber\\
& &
+\frac{1}{2}\tensor{U}{\pi_a}{\rho\rho}(\{b,c\})\left(\nabla_b\rho\right)\left(\nabla_c\rho\right)
+\frac{1}{2}\tensor{U}{\pi_a}{\rho\pi_b}(\{c,d\})\left(\nabla_c\rho\right)\left(\nabla_d\pi_b\right)\nonumber\\
& &
+\frac{1}{2}\tensor{U}{\pi_a}{\pi_b\rho}(\{c,d\})\left(\nabla_c\pi_b\right)\left(\nabla_d\rho\right)
+\frac{1}{2}\tensor{U}{\pi_a}{\pi_b\pi_c}(\{d,e\})\left(\nabla_d\pi_b\right)\left(\nabla_e\pi_c\right)+\cdots
\end{eqnarray}
Using the results for the D2Q6 model compiled in Tables~\ref{eq:sH}, \ref{eq:tH} and \ref{eq:uH}, this becomes
\begin{equation}
\mbox{\textcircled{2}}_{\pi_a}
=
-\frac{1-\sigma}{2\rho}\left(\delta_{ab}\pi_c+\delta_{bc}\pi_a-\delta_{ac}\pi_b\right)\nabla_c\pi_b+\cdots
\end{equation}
Applying incompressible scaling, we are left with the dominant terms
\begin{equation}
\boxed{
\mbox{\textcircled{2}}_{\bfpism}
=
-\frac{1-\sigma}{2\rho}\left(\bfpi\cdot\bfnabla\bfpi+\bfpi\bfnabla\cdot\bfpi-\frac{1}{2}\bfnabla\left|\bfpi\right|^2
+\calO\left(\epsilon^2\right)\right).}
\end{equation}

\section{Appendix:  Evaluation of $Q_\alpha$ for D2Q6 fluid}
\label{sec:AQ}

For a mass and momentum-conserving fluid, we wish to compute the $\rho$ and $\bfpi$ components of $Q_\alpha$, given in Eq.~(\ref{eq:QTerm}).  The $\rho$ component may be written
\begin{eqnarray}
Q_\rho
&=&
+\tensor{S}{\rho}{\rho}(\{\})\rho
+\tensor{S}{\rho}{\pi_a}(\{\})\pi_a
-\tensor{S}{\rho}{\rho}(\{a\})\nabla_a\rho
-\tensor{S}{\rho}{\pi_a}(\{b\})\nabla_b\pi_a\nonumber\\
& &
+\frac{1}{2}\tensor{S}{\rho}{\rho}(\{a,b\})\nabla_a\nabla_b\rho
+\frac{1}{2}\tensor{S}{\rho}{\pi_a}(\{b,c\})\nabla_b\nabla_c\pi_a\nonumber\\
& &
- \left(1-\sigma\right)\tensor{T}{\rho}{\rho}(\{a\})\nabla_a\rho
- \left(1-\sigma\right)\tensor{T}{\rho}{\pi_a}(\{b\})\nabla_b\pi_a\nonumber\\
& &
+ \frac{1-\sigma}{2}\tensor{T}{\rho}{\rho}(\{a,b\})\nabla_a\nabla_b\rho
+ \frac{1-\sigma}{2}\tensor{T}{\rho}{\pi_a}(\{b,c\})\nabla_b\nabla_c\pi_a\nonumber\\
& &
+ \frac{1-\sigma}{2}\tensor{U}{\rho}{\rho\rho}(\{a,b\})\left(\nabla_a\rho\right)\left(\nabla_b\rho\right)
+ \frac{1-\sigma}{2}\tensor{U}{\rho}{\rho\pi_a}(\{b,c\})\left(\nabla_b\rho\right)\left(\nabla_c\pi_a\right)\nonumber\\
& &
+ \frac{1-\sigma}{2}\tensor{U}{\rho}{\pi_a\rho}(\{b,c\})\left(\nabla_b\pi_a\right)\left(\nabla_c\rho\right)
+ \frac{1-\sigma}{2}\tensor{U}{\rho}{\pi_a\pi_b}(\{c,d\})\left(\nabla_c\pi_a\right)\left(\nabla_d\pi_b\right)\nonumber\\
& &
+ \left(1-\sigma\right)\sum_j^{b}\tensor{K}{\rho}{j}\xieq_j+\cdots
\end{eqnarray}
Using the results for the D2Q6 model compiled in Tables~\ref{eq:sH}, \ref{eq:tH} and \ref{eq:uH}, this becomes
\begin{eqnarray}
Q_\rho
&=&
+\rho
-\delta_{ab}\nabla_b\pi_a
+\frac{1}{4}\delta_{ab}\nabla_a\nabla_b\rho
+ \frac{1-\sigma}{8\rho^2}\left(\pi^2\delta_{ab}-2\pi_a\pi_b\right)\nabla_a\nabla_b\rho
\nonumber\\
& &
+ \frac{1-\sigma}{4\rho}\left(\delta_{ab}\pi_c+\delta_{ac}\pi_b-\delta_{bc}\pi_a\right)\nabla_b\nabla_c\pi_a
- \frac{1-\sigma}{4\rho^3}\left(\pi^2\delta_{ab}-\pi_a\pi_b\right)\left(\nabla_a\rho\right)\left(\nabla_b\rho\right)
\nonumber\\ & &
- \frac{1-\sigma}{4\rho^2}\left(\delta_{ab}\pi_c+\delta_{ac}\pi_b-\delta_{bc}\pi_a\right)\left(\nabla_b\rho\right)\left(\nabla_c\pi_a\right)
\nonumber\\ & &
- \frac{1-\sigma}{4\rho^2}\left(\delta_{ab}\pi_c+\delta_{ac}\pi_b-\delta_{bc}\pi_a\right)\left(\nabla_b\pi_a\right)\left(\nabla_c\rho\right)
\nonumber\\ & &
+ \frac{1-\sigma}{4\rho}\left(\delta_{ac}\delta_{bd}+\delta_{ad}\delta_{bc}
-\delta_{ab}\delta_{cd}\right)\left(\nabla_c\pi_a\right)\left(\nabla_d\pi_b\right)
+ \left(1-\sigma\right)\sum_j^{b}\tensor{K}{\rho}{j}\xieq_j+\cdots
\end{eqnarray}
Applying incompressible scaling, we are left with the dominant term
\begin{equation}
\boxed{Q_\rho = \rho+\calO\left(\epsilon^2\right).}
\end{equation}
where we have adopted incompressible scaling in the final step.

The $\bfpi$ component may be written
\begin{eqnarray}
Q_{\pi_a}
&=&
+\tensor{S}{\pi_a}{\rho}(\{\})\rho
+\tensor{S}{\pi_a}{\pi_b}(\{\})\pi_b
-\tensor{S}{\pi_a}{\rho}(\{b\})\nabla_b\rho
-\tensor{S}{\pi_a}{\pi_b}(\{c\})\nabla_c\pi_b\nonumber\\
& &
+\frac{1}{2}\tensor{S}{\pi_a}{\rho}(\{b,c\})\nabla_b\nabla_c\rho
+\frac{1}{2}\tensor{S}{\pi_a}{\pi_b}(\{c,d\})\nabla_c\nabla_d{\pi_b}\nonumber\\
& &
- \left(1-\sigma\right)\tensor{T}{\pi_a}{\rho}(\{b\})\nabla_b\rho
- \left(1-\sigma\right)\tensor{T}{\pi_a}{\pi_b}(\{c\})\nabla_c{\pi_b}\nonumber\\
& &
+ \frac{1-\sigma}{2}\tensor{T}{\pi_a}{\rho}(\{b,c\})\nabla_b\nabla_c\rho
+ \frac{1-\sigma}{2}\tensor{T}{\pi_a}{\pi_b}(2)\nabla\nabla{\pi_b}\nonumber\\
& &
+ \frac{1-\sigma}{2}\tensor{U}{\pi_a}{\rho\rho}(\{b,c\})\left(\nabla_b\rho\right)\left(\nabla_c\rho\right)
+ \frac{1-\sigma}{2}\tensor{U}{\pi_a}{\rho\pi_b}(\{c,d\})\left(\nabla_c\rho\right)\left(\nabla_d{\pi_b}\right)\nonumber\\
& &
+ \frac{1-\sigma}{2}\tensor{U}{\pi_a}{\pi_b\rho}(\{c,d\})\left(\nabla_c{\pi_b}\right)\left(\nabla_d\rho\right)
+ \frac{1-\sigma}{2}\tensor{U}{\pi_a}{\pi_b\pi_c}(\{d,e\})\left(\nabla_d{\pi_b}\right)\left(\nabla_e{\pi_c}\right)\nonumber\\
& &
+ \left(1-\sigma\right)\sum_j^{b}\tensor{K}{\pi_a}{j}\xieq_j+\cdots
\end{eqnarray}
Using the results for the D2Q6 model compiled in Tables~\ref{eq:sH}, \ref{eq:tH} and \ref{eq:uH}, this becomes
\begin{eqnarray}
Q_{\pi_a}
&=&
+\delta_{ab}\pi_b
-\frac{1}{2}\delta_{ab}\nabla_b\rho
+\frac{1}{8}
\left(\delta_{ab}\delta_{cd}+\delta_{ac}\delta_{bd}+\delta_{ad}\delta_{bc}\right)\nabla_c\nabla_d{\pi_b}\nonumber\\
& &
- \left(\frac{1-\sigma}{4\rho^2}\right)\left(\pi^2\delta_{ab}-2\pi_a\pi_b\right)\nabla_b\rho
- \left(\frac{1-\sigma}{2\rho}\right)\left(\delta_{ab}\pi_c+\delta_{bc}\pi_a-\delta_{ac}\pi_b\right)\nabla_c{\pi_b}\nonumber\\
& &
+ \left(1-\sigma\right)\sum_j^{b}\tensor{K}{\pi_a}{j}\xieq_j+\cdots
\end{eqnarray}
Applying incompressible scaling, we are left with the dominant terms
\begin{equation}
\boxed{
Q_{\bfpism}
=
\bfpi
-\frac{1}{2}\bfnabla\rho
+\frac{1}{8}\nabla^2{\bfpi}
- \left(\frac{1-\sigma}{2\rho}\right)
\left(
\bfpi\cdot\bfnabla{\bfpi}
-\frac{1}{2}\bfnabla\left|\bfpi\right|^2
\right)+\calO\left(\epsilon^2\right).}
\end{equation}

\newpage
\begin{table}[t]
\caption{The $S(N)$ tensors for the D2Q6 hydrodynamic lattice BGK model}
\begin{center}
\begin{tabular}{rc}
\toprule%
$N=0$ &
$\myvsp\tensor{S}{\rho}{\rho}(\{\}) = 1\myvsp$\\
&
$\myvsp\tensor{S}{\rho}{\pi_a}(\{\}) = 0\myvsp$\\
&
$\myvsp\tensor{S}{\pi_a}{\rho}(\{\}) = 0\myvsp$\\
&
$\myvsp\tensor{S}{\pi_a}{\pi_b}(\{\}) = \delta_{ab}\myvsp$\\
\midrule
$N=1$ &
$\myvsp\tensor{S}{\rho}{\rho}(\{a\}) = 0\myvsp$\\
&
$\myvsp\tensor{S}{\rho}{\pi_a}(\{b\}) = \delta_{ab}\myvsp$\\
&
$\myvsp\tensor{S}{\pi_a}{\rho}(\{b\}) = \frac{1}{2}\delta_{ab}\myvsp$\\
&
$\myvsp\tensor{S}{\pi_a}{\pi_b}(\{c\}) = 0\myvsp$\\
\midrule
$N=2$ &
$\myvsp\tensor{S}{\rho}{\rho}(\{a,b\}) = \frac{1}{2}\delta_{ab}\myvsp$\\
&
$\myvsp\tensor{S}{\rho}{\pi_a}(\{b,c\}) = 0\myvsp$\\
&
$\myvsp\tensor{S}{\pi_a}{\rho}(\{b,c\}) = 0\myvsp$\\
&
$\myvsp\tensor{S}{\pi_a}{\pi_b}(\{c,d\}) = \frac{1}{4}
\left(\delta_{ab}\delta_{cd}+\delta_{ac}\delta_{bd}+\delta_{ad}\delta_{bc}\right)\myvsp$\\
\midrule
$N=3$ &
$\myvsp\tensor{S}{\rho}{\rho}(\{a,b,c\}) = 0\myvsp$\\
&
$\myvsp\tensor{S}{\rho}{\pi_a}(\{b,c,d\}) = \frac{1}{4}
\left(\delta_{ab}\delta_{cd}+\delta_{ac}\delta_{bd}+\delta_{ad}\delta_{bc}\right)\myvsp$\\
&
$\myvsp\tensor{S}{\pi_a}{\rho}(\{b,c,d\}) = \frac{1}{8}
\left(\delta_{ab}\delta_{cd}+\delta_{ac}\delta_{bd}+\delta_{ad}\delta_{bc}\right)\myvsp$\\ 
&
$\myvsp\tensor{S}{\pi_a}{\pi_b}(\{c,d,e\}) = 0\myvsp$\\
\midrule
$N=4$ &
$\myvsp\tensor{S}{\rho}{\rho}(\{a,b,c,d\}) = \frac{1}{8}
\left(\delta_{ab}\delta_{cd}+\delta_{ac}\delta_{bd}+\delta_{ad}\delta_{bc}\right)\myvsp$\\
&
$\myvsp\tensor{S}{\rho}{\pi_a}(\{b,c,d,e\}) = 0\myvsp$\\
&
$\myvsp\tensor{S}{\pi_a}{\rho}(\{b,c,d,e\}) = 0\myvsp$\\
\bottomrule
\end{tabular}
\end{center}
\label{eq:sH}
\end{table}
\begin{table}[ht]
\caption{The $T(N)$ tensors for the D2Q6 hydrodynamic lattice BGK model}
\begin{center}
\begin{tabular}{rc}
\toprule%
$N=0$ &
$\myvsp\tensor{T}{\rho}{\rho}(\{\}) = 0\myvsp$\\
&
$\myvsp\tensor{T}{\rho}{\pi_a}(\{\}) = 0\myvsp$\\
&
$\myvsp\tensor{T}{\pi_a}{\rho}(\{\}) = 0\myvsp$\\
&
$\myvsp\tensor{T}{\pi_a}{\pi_b}(\{\}) = 0\myvsp$\\
\midrule
$N=1$ &
$\myvsp\tensor{T}{\rho}{\rho}(\{a\}) = 0\myvsp$\\
&
$\myvsp\tensor{T}{\rho}{\pi_a}(\{b\}) = 0\myvsp$\\
&
$\myvsp\tensor{T}{\pi_a}{\rho}(\{b\}) = \frac{1}{4\rho^2}\left(\pi^2\delta_{ab}-2\pi_a\pi_b\right)\myvsp$\\
&
$\myvsp\tensor{T}{\pi_a}{\pi_b}(\{c\}) = \frac{1}{2\rho}\left(\delta_{ab}\pi_c+\delta_{bc}\pi_a-\delta_{ac}\pi_b\right) \myvsp$\\
\midrule
$N=2$ &
$\myvsp\tensor{T}{\rho}{\rho}(\{a,b\}) = \frac{1}{4\rho^2}\left(\pi^2\delta_{ab}-2\pi_a\pi_b\right)\myvsp$\\
&
$\myvsp\tensor{T}{\rho}{\pi_a}(\{b,c\}) = \frac{1}{2\rho}\left(\delta_{ab}\pi_c+\delta_{ac}\pi_b-\delta_{bc}\pi_a\right)\myvsp$\\
&
$\myvsp\tensor{T}{\pi_a}{\rho}(\{b,c\}) = 0\myvsp$\\
&
$\myvsp\tensor{T}{\pi_a}{\pi_b}(\{c,d\}) = 0\myvsp$\\
\midrule
$N=3$ &
$\myvsp\tensor{T}{\rho}{\rho}(\{a,b\}) = 0\myvsp$\\
&
$\myvsp\tensor{T}{\rho}{\pi_a}(\{b,c\}) = 0\myvsp$\\
\bottomrule
\end{tabular}
\end{center}
\label{eq:tH}
\end{table}
\newpage
\begin{table}[ht]
\caption{The $U(N)$ tensors for the D2Q6 hydrodynamic lattice BGK model}
\begin{center}
\begin{tabular}{rc}
\toprule%
$N=0$ &
$\myvsp\tensor{U}{\rho}{\rho\rho}(\{\}) = 0\myvsp$ \\
&
$\myvsp\tensor{U}{\rho}{\rho\pi_a}(\{\}) = \tensor{U}{\rho}{\pi_a\rho}(\{\}) = 0\myvsp$\\
&
$\myvsp\tensor{U}{\rho}{\pi_a\pi_b}(\{\}) = 0\myvsp$\\
&
$\myvsp\tensor{U}{\pi_a}{\rho\rho}(\{\}) = 0\myvsp$\\
&
$\myvsp\tensor{U}{\pi_a}{\rho\pi_b}(\{\}) = \tensor{U}{\pi_a}{\pi_b\rho}(\{\}) = 0\myvsp$\\
&
$\myvsp\tensor{U}{\pi_a}{\pi_b\pi_c}(\{\}) = 0\myvsp$\\
\midrule
$N=1$ &
$\myvsp\tensor{U}{\rho}{\rho\rho}(\{ a\}) = 0\myvsp$ \\
&
$\myvsp\tensor{U}{\rho}{\rho\pi_a}(\{ b\}) = \tensor{U}{\rho}{\pi_a\rho}(\{ b\}) = 0\myvsp$\\
&
$\myvsp\tensor{U}{\rho}{\pi_a\pi_b}(\{ c\}) = 0\myvsp$\\
&
$\myvsp\tensor{U}{\pi_a}{\rho\rho}(\{ b\}) = -\frac{1}{2\rho^3}\left(\pi^2\delta_{ab}-\pi_a\pi_b\right)\myvsp$ \\
&
$\myvsp\tensor{U}{\pi_a}{\rho\pi_b}(\{ c\}) = \tensor{U}{\pi_a}{\pi_b\rho}(\{ c\}) = -\frac{1}{2\rho^2}\left(\delta_{ab}\pi_c+\delta_{bc}\pi_a-\delta_{ac}\pi_b\right)\myvsp$\\
&
$\myvsp\tensor{U}{\pi_a}{\pi_b\pi_c}(\{ d\}) = \frac{1}{2\rho}\left(\delta_{ab}\delta_{cd}+\delta_{ac}\delta_{bd}-\delta_{ad}\delta_{bc}\right)\myvsp$\\
\midrule
$N=2$ &
$\myvsp\tensor{U}{\rho}{\rho\rho}(\{a,b\}) = -\frac{1}{2\rho^3}\left(\pi^2\delta_{ab}-\pi_a\pi_b\right)\myvsp$ \\
&
$\myvsp\tensor{U}{\rho}{\rho\pi_a}(\{b,c\}) = \tensor{U}{\rho}{\pi_a\rho}(\{b,c\}) = -\frac{1}{2\rho^2}\left(\delta_{ab}\pi_c+\delta_{ac}\pi_b-\delta_{bc}\pi_a\right)\myvsp$\\
&
$\myvsp\tensor{U}{\rho}{\pi_a\pi_b}(\{c,d\}) = \frac{1}{2\rho}\left(\delta_{ac}\delta_{bd}+\delta_{ad}\delta_{bc}-\delta_{ab}\delta_{cd}\right)\myvsp$\\
&
$\myvsp\tensor{U}{\pi_a}{\rho\rho}(\{b,c\}) = 0\myvsp$ \\
&
$\myvsp\tensor{U}{\pi_a}{\rho\pi_b}(\{c,d\}) = \tensor{U}{\pi_a}{\pi_b\rho}(\{c,d\}) = 0\myvsp$\\
&
$\myvsp\tensor{U}{\pi_a}{\pi_b\pi_c}(\{d,e\}) = 0\myvsp$\\
\midrule
$N=3$ &
$\myvsp\tensor{U}{\rho}{\rho\rho}(\{a,b,c\}) = 0\myvsp$ \\
&
$\myvsp\tensor{U}{\rho}{\rho\pi_a}(\{b,c,d\}) = \tensor{U}{\rho}{\pi_a\rho}(\{b,c,d\}) = 0\myvsp$\\
&
$\myvsp\tensor{U}{\rho}{\pi_a\pi_b}(\{c,d,e\}) = 0\myvsp$\\
\bottomrule
\end{tabular}
\end{center}
\label{eq:uH}
\end{table}

\end{document}